\documentclass[amssymb, twocolumn, nobibnotes, aps, pra]{revtex4-1}

\usepackage{graphicx,epsfig,epstopdf}
\usepackage{amsmath} 
\usepackage{bm}
\usepackage{color}

\begin{document}

\title{Broadband infrared and Raman probes of excited-state vibrational molecular dynamics; Simulation protocols based on loop diagrams}

\author{Konstantin E. Dorfman}
\email{kdorfman@uci.edu}

\author{Benjamin P. Fingerhut}
\email{bfingerh@uci.edu}

\author{Shaul Mukamel}

\affiliation{Department of Chemistry, University of California, Irvine,
California 92697-2025, USA}
\date{\today}%

\begin{abstract}
Vibrational motions in electronically excited states can be observed by either time and frequency resolved infrared absorption or by off resonant stimulated Raman techniques. Multipoint correlation function expressions are derived for both signals. Three representations for the signal which suggest different simulation protocols are developed. These are based on the forward and the backward propagation of the wavefunction, sum over state expansion using an effective vibration Hamiltonian and a semiclassical treatment of a bath. We show that the effective temporal ($\Delta t$) and spectral ($\Delta\omega$) resolution  of the techniques is not controlled solely by experimental knobs but also depends on the system  dynamics  being probed. The Fourier uncertainty  $\Delta\omega\Delta t>1$ is never violated.
\end{abstract}

\maketitle

\section{Introduction}

\vspace{0.5cm}





The excited state vibrational dynamics of molecules plays a key role in many photophysical and photochemical processes and has attracted considerable experimental and theoretical attention \cite{Kukura:Science:2005,Schreier:Science:2007,Adamczyk:Science:2009,Kuramochi:JPCLett:2012}.   Real time structural information about rearrangement of atoms in complex reactions can be inferred directly from time resolved vibrational spectroscopy \cite{Kukura:AnnurevPhysChem:2007,Haiser:AngewChem:2011,Fingerhut:NJP:2012}. Typically  an ultrashort  laser pulse in the visible or the UV excites the molecule to a bright valence excited state, launching a photoreaction or non-adiabatic relaxation process. The vibrational dynamics can then  be probed either by the absorption of a delayed  IR probe pulse \cite{Fayerbook:2013,Anfinrud_PNAS:89,Sawyer:JPCA:2008,Keane:JACS:2011,Lynch:JPCA:2012,Bingaman:JPCA:2012} or by a spontaneous or stimulated Raman process \cite{Schrader:CPL:2004}.

Unique marker bands in UV pump/IR probe signals serve as fingerprint of the excited state evolution allowing to resolve transient reaction \cite{Schreier:Science:2007,Mohammed:Science:2005}, intermediates structural details \cite{Lim:Science:1995} and reveal the reaction mechanism. Such investigations helped identify the real time  reaction mechanism leading to the formation  of photolesions in DNA  nucleobases \cite{Schreier:Science:2007, Schreier:JACS:2009,Haiser:AngewChem:2011,Fingerhut:NJP:2012}, to monitor isomerization reactions in protein environments \cite{Heyne:JACS:2005}, to resolve the consecutive steps in proton transfer reactions \cite{Mohammed:Science:2005}, to identify
the participating ion pairs upon photoinduced bimolecular electron transfer \cite{Mohammed:AngewChem:2008}
and to follow light-induced electrocyclic reactions \cite{Mohammed:JPCA:2009,Nibbering:AnnRevPhysChem:2005}.  Frequency shifts of IR marker bands have also been used to monitor the response of the local environment \cite{Szyc:AngewChem:2010,Mohammed:JPCA:2011,Premont-Schwarz:JPCA:2011,Xiao:JPCA:2012} and molecular  energy redistribution  \cite{Kozich:JCP:2002}.
More elaborate pulse sequences allow to spread the IR signal in two dimensions, resolving the couplings between localized vibrations \cite{Hamm:PNAS:1999,Zanni:PNAS:2001,Zhu09}.

Spontaneous Raman \cite{Miz97} has long been used as an alternative probe. Recent stimulated Raman measurements that employ a femtosecond and a picosecond pulse had generated considerable excitement \cite{McCamant:JPCA:2003,Lee:JCP:2004, Laimgruber:AngewChem:2005, Kukura:AnnurevPhysChem:2007,Umapathy:JRS:2009,Mehlenbacher:JCP:2009,Takeuchi:Science:2008,Kuramochi:JPCLett:2012,Xin02,Mul02,Kee04,Vol05,Vac07}. A rich pattern of narrow (10 cm$^{-1}$) vibrational lines has been reported in 25 fs intervals. Applications were made to $p$NA, $p$DNA \cite{An:OptCommun:2002}, rhodopsin \cite{Kukura:Science:2005}, carbon dioxide \cite{Roy:JRS:2010}, bacterial endospores \cite{Pestov:Science:2007}, and other systems. Frequency domain stimulated Raman has proven validity in cell imaging \cite{Zumbusch:PhysRevLett:99}.

 
In this paper we focus on two techniques, both starting with an optical pump pulse but followed  by a different detection: frequency - dispersed broadband infrared probe (FDIR) or off resonant stimulated Raman spectroscopy (SRS) \cite{Kukura:AnnurevPhysChem:2007}. We show how both techniques can be described and interpreted with minor modifications using very similar vibrational correlation functions . The signals are intuitively described by  loop diagrams which connect them to forward and backward propagation of the wavefunction.

We present a general analysis and derive closed expressions that can be used for microscopic quantum simulations of both infrared and Raman signals. Three representations for these correlation functions are presented each suggesting a different simulation strategy. The first is based on the numerical propagation of the wavefunction which includes all relevant electronic and nuclear (including bath) degrees of freedom explicitly. This is the most general, expensive and accurate method \cite{Kosloff:annurevpc:1994,Chen:JCP:2006}. A  second protocol  uses  a Sum Over States (SOS) expansion  of the signals. Here we must diagonalize an effective  vibrational hamiltonian. This offers a numerically more tractable  algorithm when it is  possible to truncate the relevant phase space. The third approach is  semiclassical. A small vibrational system is treated quantum mechanically and a  classical bath which causes a time dependent modulation of the system Hamiltonian is added. This is the simplest theory to implement by e.g. assuming that the vibrational frequencies change with  time. This change can be either introduced phenomenologically or  by using atomistic molecular dynamics simulations.

The time and frequency in these experiments are controlled by independent  knobs.   We can formally define uncertainties $\Delta t$ and $\Delta\omega$ associated with the pulse duration and the frequency resolution of a spectrometer. This suggests that there is no lower bound to the product $\Delta\omega\Delta t$; the measurement can apparently be interpreted in terms of instantaneous snapshots with high spectral resolution. For example recent experiments \cite{Kukura:Science:2005,Kukura:AnnurevPhysChem:2007} use pulses ($<50$ fs) and reported spectral features ($<10$ cm$^{-1}$) such that $\Delta\omega\Delta t\sim 0.5$ ps cm$^{-1}$ which is an order of magnitude smaller than the Fourier uncertainty for Gaussian pulses. An additional goal of this paper is to provide a proper definition of $\Delta\omega$ and $\Delta t$ and show that they are not purely instrumental but depend on the system as well. We find that the simple snapshot interpretation is false. We discuss the limitations of the spectral and temporal  resolutions of these techniques and how they can be manipulated. 

  \begin{figure}[t]
\begin{center}
\includegraphics[trim=0cm 0cm 0cm 0cm,angle=0, width=0.45\textwidth]{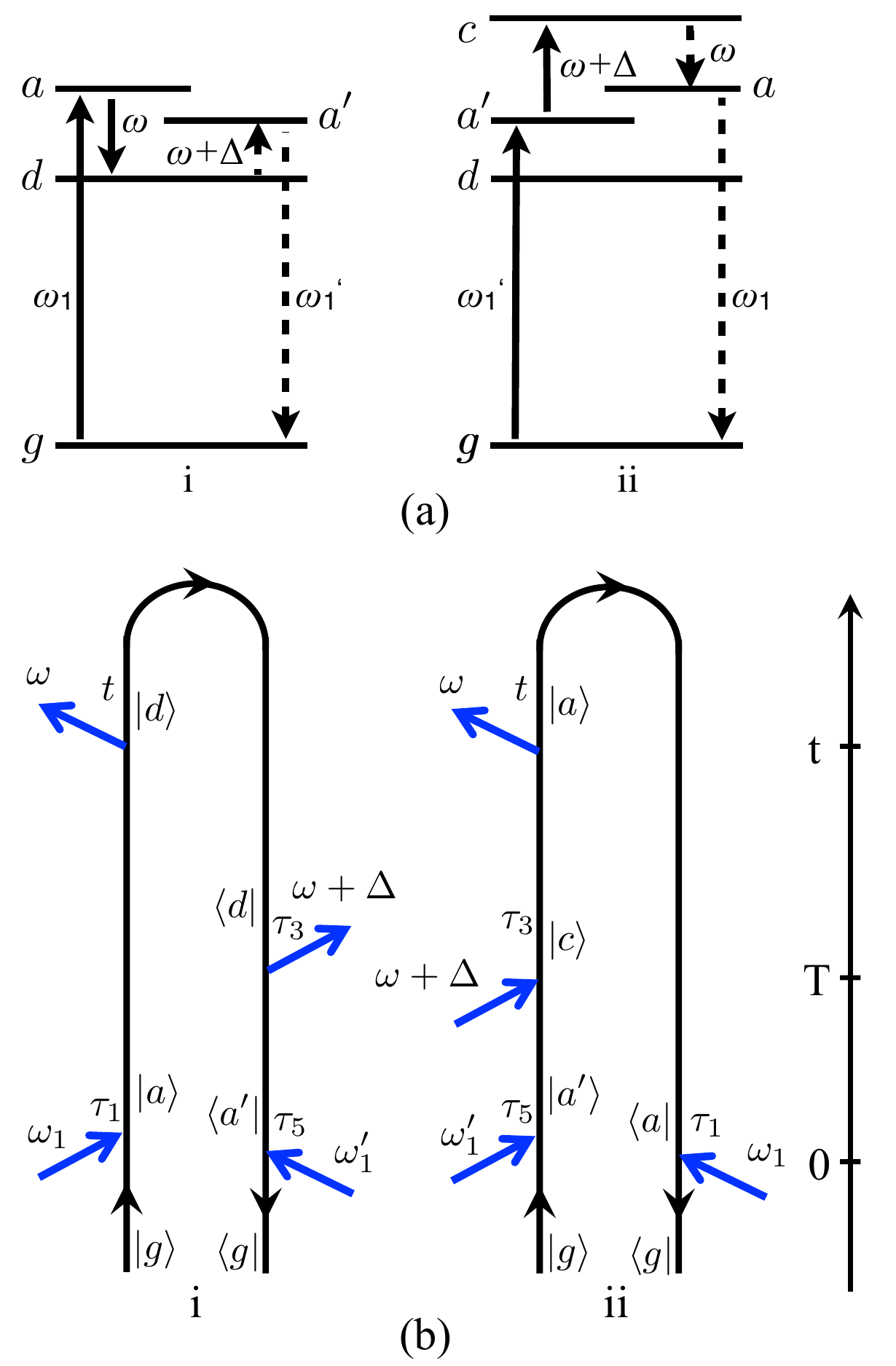}
\end{center}
\caption{(Color online) FDIR: Level scheme - (a), closed-time path-loop diagrams - (b).}
\label{fig:IR}
\end{figure}

\section{Loop diagram representation of frequency-dispersed stimulated signals}

Stimulated optical signals are defined as the energy change of the electromagnetic field
\begin{equation}
S=\int_{-\infty}^{\infty}\frac{d}{dt}\langle \mathcal{E}^{\dagger}(t)\mathcal{E}(t)\rangle dt.
\end{equation}
 The radiation-matter interaction Hamiltonian in the rotating wave approximation (RWA) is
\begin{equation}\label{eq:H1}
H'(t)=V(t)\mathcal{E}^{\dagger}(t)+H.c.,
\end{equation}
where $V(t)+V^{\dagger}(t)$ is a Heisenberg dipole operator and the electric field operator $E(t)=\mathcal{E}(t)+\mathcal{E}^{\dagger}(t)$. Both are separated into positive (non dagger) and negative (dagger) frequency components (lowering and raising photon operators, respectively). The dipole operator  is  given by the sum of the electronic and nuclear dipole moments $V(t)=V_e(t)+V_n(t)$. 

The Heisenberg equation of motion for the field operator $E(t)$ then gives for the above integrated signal
\begin{align}\label{eq:S0}
S&=\frac{2}{\hbar}\int_{-\infty}^{\infty}dt'~\mathcal{I}\langle P(t')\mathcal{E}^{\dagger}(t')\rangle\notag\\
&=\frac{2}{\hbar}\int_{-\infty}^{\infty}\frac{d\omega'}{2\pi}\mathcal{I}\langle P(\omega')\mathcal{E}^{\dagger}(\omega')\rangle,
\end{align}
where where $\mathcal{I}$ denotes the imaginary part,
\begin{equation}
P(\omega)=\int_{-\infty}^{\infty}dt P(t)e^{i\omega t}
\end{equation}
with $P(t)=\langle V(t)\rangle$ representing the nonlinear polarization that arise from the interaction with the pump and the probe pulses. The angular brackets denote $\langle...\rangle=\text{Tr}[\rho(t)...]$ with the density operator $\rho(t)$ defined in the joint field-matter space of
the entire system. In practice, the temporal or spectral range of the integrations in Eq. (\ref{eq:S0}) is restricted by the response function of the detector. If the detector contains a narrow time gate with nearly $\delta$ function response $\delta(t'-t)$, Eq. (\ref{eq:S0}) yields
\begin{equation}\label{eq:St}
S_{TG}(t;\Gamma)=\frac{2}{\hbar}\mathcal{I}\langle P(t)\mathcal{E}^{\dagger}(t)\rangle,
\end{equation}
where $\Gamma$ denotes  a set of parameters that characterize the various laser pulses. Similarly if the detector consists of a spectrometer with narrow frequency response $\delta(\omega'-\omega)$, we obtain the frequency-gated signal
\begin{equation}\label{eq:Sw}
S_{FG}(\omega;\Gamma)=\frac{2}{\hbar}\mathcal{I}\langle P(\omega)\mathcal{E}^{\dagger}(\omega)\rangle,
\end{equation}
Note that the two signals Eqs. (\ref{eq:St}) and (\ref{eq:Sw}) carry different information and are not related by a simple Fourier transform. A Wigner spectrogram representation \cite{Sto94,Dor12,Dor121} was used in \cite{Pol10} for the integrated pump probe signals Eq. (\ref{eq:S0}). Here we use loop diagrams to describe the more detailed frequency-  or time-gated signals  (\ref{eq:Sw}) and (\ref{eq:St}), respectively. For clarity in the following we focus on the frequency-gated expressions, the corresponding time-gated signals are given in the Appendix \ref{app:time}.

  \begin{figure}[t]
\begin{center}
\includegraphics[trim=0cm 0cm 0cm 0cm,angle=0, width=0.45\textwidth]{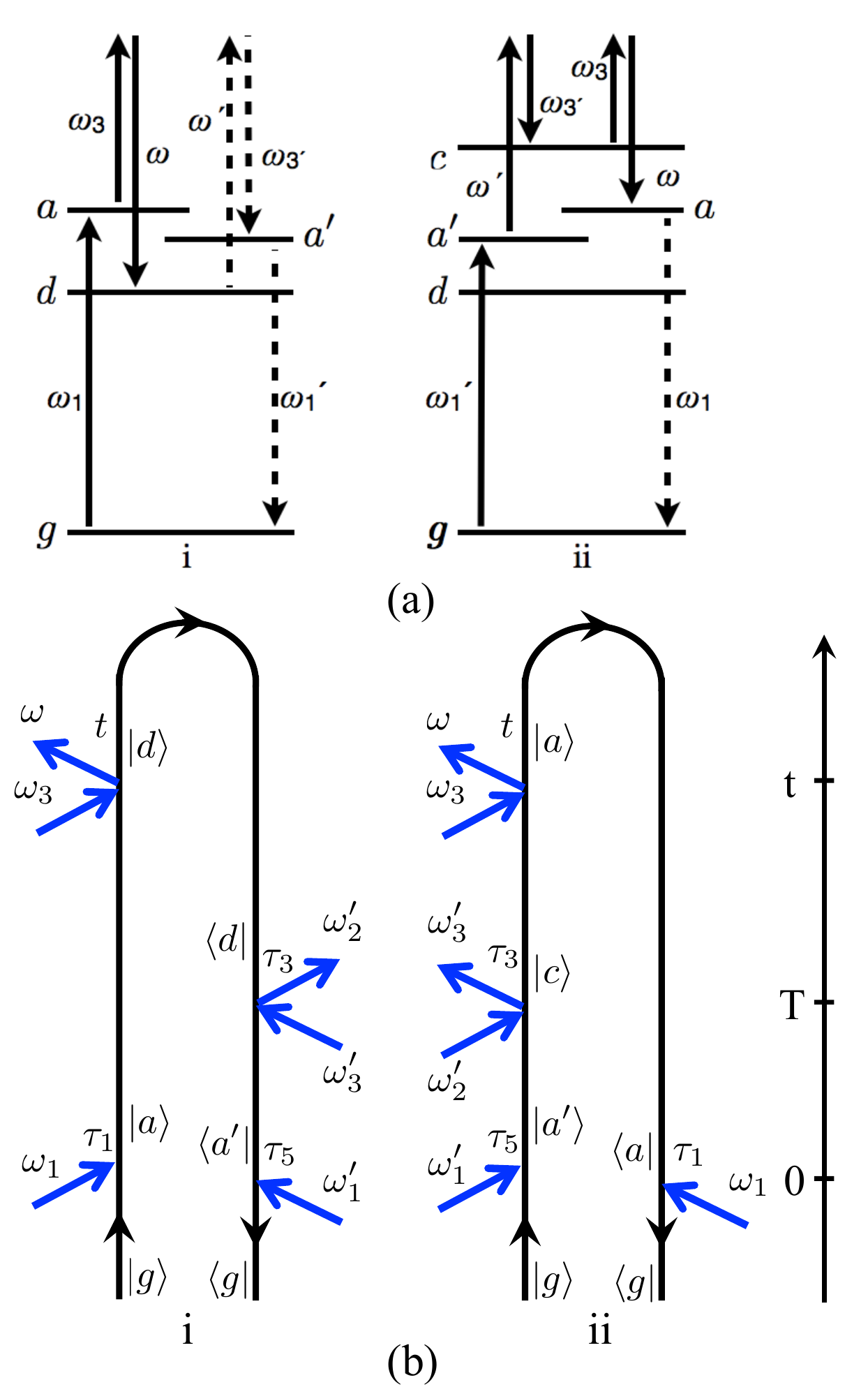}
\end{center}
\caption{(Color online) SRS: Level scheme - (a), closed-time path-loop diagrams - (b).}
\label{fig:Ram}
\end{figure}

\section{First protocol; Numerical propagation of the wave function}

We start with the visible-pump/IR-probe signal as sketched in Fig. \ref{fig:IR}a. FDIR is somewhat simpler than SRS since it only involves four rather than six radiation-matter interactions. The pump pulse centered at time $\tau_3=0$ promotes the system from its ground electronic state $g$ to the vibrational state $a$ of an excited electronic state  and launches the vibrational dynamics. The IR probe pulse centered around $\tau_3=T$ can then either stimulate emission that couples the vibrational state $a$ and lower vibrational state $d$ or an absorption to higher vibrational state $c$. The signal is defined as the change in probe intensity and is either time-gated (Eq. (\ref{eq:St}) or frequency-gated (Eq. (\ref{eq:Sw})). Both can be represented by the loop diagrams shown in Fig. \ref{fig:IR}b which contain the four field-matter interactions -two with each pulse. Diagram rules are given in Ref. \cite{Rah10}. Note that the signals (\ref{eq:St}) and (\ref{eq:Sw}) are expressed in terms of $\mathcal{E}^{\dagger}$ which makes the arrow corresponding the the last interaction pointing to the left. We further choose the last interaction to occur on the left branch. This choice removes any ambiguity in the diagrams rules without loss of generality.

The electric field operator consists of the pump field $1$ and a probe field $2$

\begin{equation}
\mathcal{E}(t)=\mathcal{E}_1(t)+\mathcal{E}_2(t-T),
\end{equation}
where $T$ represents the delay of the probe pulse relative to the pump. The signal  is given by the two loop diagrams shown in Fig. \ref{fig:IR}b, plus their complex conjugates. These give for the frequency-gated signal (\ref{eq:Sw}) 
\begin{align}\label{eq:Swb}
S_{IR}(\omega,T)=\mathcal{I}\int_{-\infty}^{\infty}\frac{d\Delta}{2\pi}\mathcal{E}_2^{*}(\omega)\mathcal{E}_2(\omega+\Delta)\tilde{S}_{IR}(\omega,T;\Delta),
\end{align}
where the $\Delta$-dispersed signal is given by the two diagrams $\tilde{S}_{IR}(\omega,T;\Delta)=\tilde{S}_{IR}^{(i)}(\omega,T;\Delta)+\tilde{S}_{IR}^{(ii)}(\omega,T;\Delta)$ and
\begin{widetext}
\begin{align}\label{eq:Si}
\tilde{S}_{IR}^{(i)}(\omega,T;\Delta)=&\frac{2}{\hbar}\int_{-\infty}^{\infty}dt\int_{-\infty}^td\tau_1\int_{-\infty}^td\tau_3\int_{-\infty}^{\tau_3}d\tau_5\mathcal{E}_1^{*}(\tau_5)\mathcal{E}_1(\tau_1)\notag\\
&\times\langle V_eG^{\dagger}(\tau_3,\tau_5)V_n^{\dagger}G^{\dagger}(t,\tau_3)V_nG(t,\tau_1)V_e^{\dagger}\rangle e^{i\omega(t-\tau_3)-i\Delta(\tau_3-T)},
\end{align}
\begin{align}\label{eq:Sii}
\tilde{S}_{IR}^{(ii)}(\omega,T;\Delta)=&\frac{2}{\hbar}\int_{-\infty}^{\infty}dt\int_{-\infty}^td\tau_1\int_{-\infty}^td\tau_3\int_{-\infty}^{\tau_3}d\tau_5\mathcal{E}_1(\tau_5)\mathcal{E}_1^{*}(\tau_1)\notag\\
&\times\langle V_eG^{\dagger}(t,\tau_1)V_nG(t,\tau_3)V_n^{\dagger}G(\tau_3,\tau_5)V_e^{\dagger}\rangle e^{i\omega(t-\tau_3)-i\Delta(\tau_3-T)}.
\end{align}
\end{widetext}
$\tilde{S}(\omega,T;\Delta)$ represents the contribution of the $\omega$ and $\omega+\Delta$ frequency components of $\mathcal{E}_2$ to the signal, where $\omega$ is the detected frequency. The signal is obtained by integration over $\Delta$. Here $G(t_1,t_2)=(-i/\hbar)\theta(t_1-t_2)e^{-iH(t_1-t_2)}$ is the retarded Green's function. Even though this Green's function only depends on the  difference of its two time arguments, we retain both arguments and write $G(t_1,t_2)$ rather than $G(t_1-t_2)$. This is done since in the reduced (semiclassical) description to be developed later when the system is coupled to some stochastic bath degrees of freedom, time translational invariance is lost and $G$ then depends on both arguments. The corresponding time-gated signal (\ref{eq:St}) is given by Eqs. (\ref{eq:PB}) - (\ref{eq:PC}). Eqs. (\ref{eq:Si}) - (\ref{eq:Sii}) may be  simplified further when the first pulse is impulsive. We can then set $\mathcal{E}_1(\tau)=\mathcal{E}_1\delta(\tau)$ and the $\tau_1$ and $\tau_5$ integrations can be eliminated.

Diagram ($i$) (Eq. (\ref{eq:Si})) can be understood using a forward and backward time evolving vibronic wave packet. First, the pulse $\mathcal{E}_1$ electronically excites the molecule via $V_e^{\dagger}$. The wavefunction then propagates  forward in time from $\tau_1$ to $t$. Then the IR probe pulse $\mathcal{E}_2$ deexcites the vibrational transition to the lower vibrational level via $V_n$ which then propagates backward in time from $t$ to $\tau_3$. Pulse $\mathcal{E}_2$ excites the vibration via $V_n^{\dagger}$ and the wavefunction propagates backward in time from $\tau_3$ to $\tau_5$. The final deexcitation by pulse $\mathcal{E}_1$ returns the system to its initial state by acting with $V_e$. Diagram $ii$ Eq. (\ref{eq:Sii}) can be interpreted similarly. Following initial electronic excitation the wavefunction propagates forward in time from $\tau_5$ to $\tau_3$. At this point a vibrational excitation promotes it to the higher vibrational state and the wavefunction propagates forward in time from $\tau_3$ to $t$. After vibrational deexcitation it then propagates backward from $t$ to $\tau_1$ where an electronic excitation brings the system back in its initial ground state.

In Eqs. (\ref{eq:Si}) - (\ref{eq:Sii}) the matter correlation function is given in the time domain. Alternatively one can read the signal (\ref{eq:Swb}) from the diagrams when both field and matter correlation functions are given in the frequency domain
\begin{widetext}
\begin{align}\label{eq:Siw}
S_{IR}^{(i)}(\omega,T)=\mathcal{I}\frac{4\pi}{\hbar}\int_{-\infty}^{\infty}\frac{d\omega'}{2\pi}\frac{d\omega_1}{2\pi}\frac{d\omega_1'}{2\pi}&\mathcal{E}_2^{*}(\omega)\mathcal{E}_2(\omega')\mathcal{E}_1^{*}(\omega_1)\mathcal{E}_1(\omega_1')\delta(\omega-\omega'+\omega_1-\omega_1')\notag\\
\times&\langle V_eG^{\dagger}(\omega_g+\omega_1)V_n^{\dagger}G^{\dagger}(\omega_g+\omega_1-\omega')V_nG(\omega_g+\omega_1')V_e^{\dagger}\rangle,
\end{align}
\begin{align}\label{eq:Siiw}
S_{IR}^{(ii)}(\omega,T)=\mathcal{I}\frac{4\pi}{\hbar}\int_{-\infty}^{\infty}\frac{d\omega'}{2\pi}\frac{d\omega_1}{2\pi}\frac{d\omega_1'}{2\pi}&\mathcal{E}_2^{*}(\omega)\mathcal{E}_2(\omega')\mathcal{E}_1^{*}(\omega_1)\mathcal{E}_1(\omega_1')\delta(\omega-\omega'+\omega_1-\omega_1')\notag\\
\times&\langle V_eG^{\dagger}(\omega_g+\omega_1)V_nG(\omega_g+\omega'+\omega_1')V_n^{\dagger}G(\omega_g+\omega_1')V_e^{\dagger}\rangle.
\end{align}
\end{widetext}
Here $G(\omega)=h^{-1}/[\omega-H/\hbar+i\epsilon]$, $\delta(\omega-\omega'+\omega_1-\omega_1')$ represents the energy conservation that follows from time translation symmetry of all four field-matter interactions. One can separate the preparation pulse $\mathcal{E}_1$ and break the $\delta$-function as follows
\begin{align}\label{eq:bdf}
&\delta(\omega-\omega'+\omega_1-\omega_1')\notag\\
&=\int_{-\infty}^{\infty}d\Delta\delta(\omega-\omega'+\Delta)\delta(\omega_1-\omega_1'-\Delta),
\end{align}
where $\Delta$ defines the spectral bandwidth of the incoming pulse which translates into the spectral bandwidth of the relevant matter degrees of freedom. Eqs. (\ref{eq:Siw}) - (\ref{eq:Siiw}) then yield
\begin{widetext}
\begin{align}\label{eq:Siww}
S_{IR}^{(i)}(\omega,T)=\mathcal{I}\frac{2}{\hbar}\int_{-\infty}^{\infty}d\Delta d\omega_1&\mathcal{E}_2^{*}(\omega)\mathcal{E}_2(\omega+\Delta)\mathcal{E}_1^{*}(\omega_1)\mathcal{E}_1(\omega_1-\Delta)\notag\\
\times&\langle V_eG^{\dagger}(\omega_g+\omega_1)V_n^{\dagger}G^{\dagger}(\omega_g+\omega_1-\omega-\Delta)V_nG(\omega_g+\omega_1-\Delta)V_e^{\dagger}\rangle,
\end{align}
\begin{align}\label{eq:Siiww}
S_{IR}^{(ii)}(\omega,T)=\mathcal{I}\frac{2}{\hbar}\int_{-\infty}^{\infty}d\Delta d\omega_1&\mathcal{E}_2^{*}(\omega)\mathcal{E}_2(\omega+\Delta)\mathcal{E}_1^{*}(\omega_1)\mathcal{E}_1(\omega_1-\Delta)\notag\\
\times&\langle V_eG^{\dagger}(\omega_g+\omega_1)V_nG(\omega_g+\omega+\omega_1)V_n^{\dagger}G(\omega_g+\omega_1-\Delta)V_e^{\dagger}\rangle.
\end{align}
\end{widetext}

We now turn to the electronically off-resonant SRS signal shown in Fig. \ref{fig:Ram}a,b, which is completely analogous to the FDIR signal. Even though these signals represent different physical processes and even involve different number of field-matter interactions they can be described using very similar diagrams where we simply replace $V_n\to\alpha_n$ and $\omega$ by $\omega-\omega_3$. In SRS the pump pulse initiates the vibrational dynamics in the  excited electronic state. Pulse $3$ and the probe then induce the Raman process (see. Fig. \ref{fig:Ram}a). The relevant diagrams are shown in Fig. \ref{fig:Ram}b (plus their complex conjugates). An electronically off-resonant Raman process induced by pulses $2$ and $3$ is instantaneous  since by Heisenberg uncertainty the system can only spend a very short time in the intermediate state. The Raman process is thus described by an effective field/matter interaction Hamiltonian
\begin{equation}\label{eq:Hp}
H'(t)=\alpha_n\mathcal{E}_2^{\dagger}(t)\mathcal{E}_3(t)+\mathcal{E}_1^{\dagger}(t)V_e(t)+H.c.,
\end{equation}
where $\alpha_n=\tilde{\alpha}_n+\tilde{\alpha}_n^{\dagger}$ is the excited state polarizability that couples fields $2$ and $3$ parametrically via a Raman process. It is symmetric (real) operator. The time-domain signal (\ref{eq:St}) can be read directly from diagrams ($i$) and ($ii$). Assuming that pulse $3$ is narrow band (picosecond) and set $\mathcal{E}_3(t-T)=\mathcal{E}_3e^{-i\omega_3(t-T)}$. We obtain the frequency gated Raman analogues of Eqs. (\ref{eq:Si}) - (\ref{eq:Sii})

\begin{widetext}
\begin{align}\label{eq:Sisr}
\tilde{S}_{SRS}^{(i)}(\omega-\omega_3,T;\Delta)=&\frac{2}{\hbar}\int_{-\infty}^{\infty}dt\int_{-\infty}^td\tau_1\int_{-\infty}^td\tau_3\int_{-\infty}^{\tau_3}d\tau_5 |\mathcal{E}_3|^2\mathcal{E}_1^{*}(\tau_5)\mathcal{E}_1(\tau_1)\notag\\
&\times\langle V_eG^{\dagger}(\tau_3,\tau_5)\alpha_nG^{\dagger}(t,\tau_3)\alpha_nG(t,\tau_1)V_e^{\dagger}\rangle e^{i(\omega-\omega_3)(t-\tau_3)-i\Delta(\tau_3-T)},
\end{align}
\begin{align}\label{eq:Siisr}
\tilde{S}_{SRS}^{(ii)}(\omega-\omega_3,T;\Delta)=&\frac{2}{\hbar}\int_{-\infty}^{\infty}dt\int_{-\infty}^td\tau_1\int_{-\infty}^td\tau_3\int_{-\infty}^{\tau_3}d\tau_5|\mathcal{E}_3|^2\mathcal{E}_1(\tau_5)\mathcal{E}_1^{*}(\tau_1)\notag\\
&\times\langle V_eG^{\dagger}(t,\tau_1)\alpha_nG(t,\tau_3)\alpha_nG(\tau_3,\tau_5)V_e^{\dagger}\rangle e^{i(\omega-\omega_3)(t-\tau_3)-i\Delta(\tau_3-T)}.
\end{align}
The corresponding time-gated signals are given by Eqs. (\ref{eq:Pisr}) - (\ref{eq:Piisr}). Similarly to Eqs. (\ref{eq:Siw}) - (\ref{eq:Siiw}) we can recast (\ref{eq:Sisr}) - (\ref{eq:Siisr}) using frequency domain matter correlation functions
\begin{align}\label{eq:Siwsr}
S_{SRS}^{(i)}(\omega-\omega_3,T)=\mathcal{I}\frac{4\pi}{\hbar}\int_{-\infty}^{\infty}\frac{d\omega'}{2\pi}\frac{d\omega_1}{2\pi}\frac{d\omega_1'}{2\pi}&|\mathcal{E}_3|^2\mathcal{E}_2^{*}(\omega)\mathcal{E}_2(\omega')\mathcal{E}_1^{*}(\omega_1)\mathcal{E}_1(\omega_1')\delta(\omega-\omega'+\omega_1-\omega_1')\notag\\
\times&\langle V_eG^{\dagger}(\omega_g+\omega_1)\alpha_nG^{\dagger}(\omega_g+\omega_1-\omega'+\omega_3)\alpha_nG(\omega_g+\omega_1')V_e^{\dagger}\rangle,
\end{align}
\begin{align}\label{eq:Siiwsr}
S_{SRS}^{(ii)}(\omega-\omega_3,T)=\mathcal{I}\frac{4\pi}{\hbar}\int_{-\infty}^{\infty}\frac{d\omega'}{2\pi}\frac{d\omega_1}{2\pi}\frac{d\omega_1'}{2\pi}&|\mathcal{E}_3|^2\mathcal{E}_2^{*}(\omega)\mathcal{E}_2(\omega')\mathcal{E}_1^{*}(\omega_1)\mathcal{E}_1(\omega_1')\delta(\omega-\omega'+\omega_1-\omega_1')\notag\\
\times&\langle V_eG^{\dagger}(\omega_g+\omega_1)\alpha_nG(\omega_g+\omega'+\omega_1'-\omega_3)\alpha_nG(\omega_g+\omega_1')V_e^{\dagger}\rangle.
\end{align}
Breaking up the $\delta$-function according to Eq. (\ref{eq:bdf}) we get 
\begin{align}\label{eq:Siwwsr}
S_{SRS}^{(i)}(\omega-\omega_3,T)=\mathcal{I}\frac{2}{\hbar}\int_{-\infty}^{\infty}d\Delta d\omega_1&|\mathcal{E}_3|^2\mathcal{E}_2^{*}(\omega)\mathcal{E}_2(\omega+\Delta)\mathcal{E}_1^{*}(\omega_1)\mathcal{E}_1(\omega_1-\Delta)\notag\\
\times&\langle V_eG^{\dagger}(\omega_g+\omega_1)\alpha_nG^{\dagger}(\omega_g+\omega_1-\omega+\omega_3-\Delta)\alpha_nG(\omega_g+\omega_1-\Delta)V_e^{\dagger}\rangle,
\end{align}
\begin{align}\label{eq:Siiwwsr}
S_{SRS}^{(ii)}(\omega-\omega_3,T)=\mathcal{I}\frac{2}{\hbar}\int_{-\infty}^{\infty}d\Delta d\omega_1&|\mathcal{E}_3|^2\mathcal{E}_2^{*}(\omega)\mathcal{E}_2(\omega+\Delta)\mathcal{E}_1^{*}(\omega_1)\mathcal{E}_1(\omega_1-\Delta)\notag\\
\times&\langle V_eG^{\dagger}(\omega_g+\omega_1)\alpha_nG(\omega_g+\omega-\omega_3+\omega_1)\alpha_nG(\omega_g+\omega_1-\Delta)V_e^{\dagger}\rangle.
\end{align}

\end{widetext}

The simulation protocol based on these equations requires the full forward ($G$) and backward  ($G^{\dagger}$) propagation of the wavefunction retaining all electronic and nuclear degrees of freedom. This task can be accomplished by numerically exact propagation techniques, based on the split-operator Fourier-transform,  the  short iterative Lanczos method or a Chebyshev expansion \cite{Leforestier:JComputPhys:1991,Kosloff:annurevpc:1994}, where the wavefunction is commonly  expanded in the set of orthogonal eigenstates of $H$. Non-adiabatic effects can be conviniently be accounted for either in a diabtic or adiabatic basis of the participating electronic states \cite{Tannor:Book:2006}.
The major drawback of this numerically exact treatment is that the  computational effort and storage requirements grow exponentially with the number of degrees of freedom considered which limits their application to molecular systems with less than six degrees of freedom (4 atoms). 
The change to a nonorthogonal representations  of the time-dependent wavefuntion  allows to evaluate the Trotter expansion analytically and thus to avoid the unfavourable scaling  behaviour
which is accordingly not an intrinsic property  of the powerful propagators \cite{Wu:JCP:2004,Chen:JCP:2006}.
The approximate  multiconÞguration time-dependent Hartree (MCTDH) method \cite{Beck:PhysRep:2000} formally still scales exponentially but superior scaling and less memory requirements compared to the exact propagation  methods can be achieved if the number of degrees of freedom and contraction coefficients  is large. 
A major drawback of all propagation methods is that the global multi-dimensional potential energy surface has to be known \emph{a priori}. Approximate direct quantum dynamical approaches like e.g. the variational multi-configuration Gaussian wavepacket method \cite{Lasorne:CPL:2006} or \emph{ab initio} multiple spawning \cite{BenNun:JPC:1996,BenNun:JCP:1998} which rely on Gaussian functions as basis set circumvent this shortcoming as the potential energy surface is only sampled in the actual fraction of space where it is actually required.
In some applications it may be desirable to only consider  a few vibrational modes explicitly and treat the rest classically. Even in this case we may use the Green's functions expressions (\ref{eq:Si}) - (\ref{eq:Sii}) propagated forward and backward along the loop under an effective time dependent Hamiltonian \cite{Falvo:JCP:2009}.

\section{Second protocol; Sum Over States expansion}

One can evaluate the matter correlation functions in Eqs. (\ref{eq:Si}) - (\ref{eq:Sii}) by expanding them in the eigenstates of the total system. Again in this approach all bath degrees of freedom cannot be separated and must be included explicitly. The resulting SOS expansion provides useful insights and a convenient computational algorithm. Starting with Eqs. (\ref{eq:Si}) - (\ref{eq:Sii}). A frequency-gated signal can be expressed

\begin{align}\label{eq:Siwir}
&S_{IR}^{(i)}(\omega,T)=-\mathcal{I}\frac{2i}{\hbar^4}\sum_{a,a',d}\frac{\mu_{ga'}\mu_{ag}^{*}\mu_{a'd}^{*}\mu_{ad}e^{-(i\omega_{aa'}+\gamma_{aa'})T}}{\omega-\omega_{ad}+i\gamma_{ad}}\notag\\
&\times\mathcal{E}_2^{*}(\omega)\mathcal{E}_2(\omega-\omega_{aa'}+i\gamma_{aa'})\mathcal{E}_1^{*}(\omega_{a'}+i\gamma_{a'})\mathcal{E}_1(\omega_a-i\gamma_a)
\end{align}
\begin{align}\label{eq:Siiwir}
&S_{IR}^{(ii)}(\omega,T)=-\mathcal{I}\frac{2i}{\hbar^4}\sum_{a,a',c}\frac{\mu_{ga'}\mu_{ag}^{*}\mu_{a'c}^{*}\mu_{ac}e^{-(i\omega_{a'a}+\gamma_{a'a})T}}{\omega-\omega_{a'c}+i\gamma_{a'c}}\notag\\
&\times\mathcal{E}_2^{*}(\omega)\mathcal{E}_2(\omega-\omega_{a'a}+i\gamma_{a'a})\mathcal{E}_1^{*}(\omega_{a'}+i\gamma_{a'})\mathcal{E}_1(\omega_a-i\gamma_a).
\end{align}
The corresponding time-gated signals are given by Eqs. (\ref{eq:Sitir}) - (\ref{eq:Siitir}). 

So far we had expanded the density operator starting with the ground state and including the preparation pulse $\mathcal{E}_1$. Alternatively we can avoid the explicit treatment of preparation and simply assume that the system has been initially prepared in  non stationary state represented by the density operator $\rho_{aa'}$. Elaborate pulse sequences can be used in this preparation. We can then evaluate the matter correlation function that corresponds to the last two interactions with the probe pulse for the $i$ contribution:

\begin{align}
&\langle\hat{V}_{nL}(t)\hat{V}_{nR}^{\dagger}(\tau_3)\rangle=Tr\left[\hat{V}_n^{\dagger}(\tau_3)\rho\hat{V}_n(t)\right]\notag\\
&=\sum_{a,a',d}\rho_{a,a'}\langle d|\hat{V}_n(t)|a\rangle\langle a'|\hat{V}_n^{\dagger}(\tau_3)|d\rangle\notag\\
&=\sum_{a,a',d}\rho_{aa'}\mu_{ad}\mu^*_{a'd}e^{-[i\omega_{ad}+\gamma_{ad}]t}e^{[i\omega_{a'd}+\gamma_d-\gamma_{a'}]\tau_3},
\end{align}
The $\Delta$ - dispersed signal (\ref{eq:Si}) - (\ref{eq:Sii}) then yield

\begin{align}\label{eq:Siwir1}
&\tilde{S}_{IR}^{(i)}(\omega,T;\Delta)=\notag\\
&-\frac{4\pi i}{\hbar^2}\sum_{a,a',d}\frac{\mu_{a'd}^{*}\mu_{ad}\rho_{aa'}\delta(\Delta-\omega_{a'a}+i\gamma_{a'a})e^{i\Delta T}}{\omega+\Delta-\omega_{a'd}+i(\gamma_{d}-\gamma_{a'})},
\end{align}

\begin{align}\label{eq:Siiwir1}
&\tilde{S}_{IR}^{(ii)}(\omega,T;\Delta)=\notag\\
&-\frac{4\pi i}{\hbar^2}\sum_{a,a',c}\frac{\mu_{a'c}^{*}\mu_{ac}\rho_{a'a}\delta(\Delta-\omega_{aa'}-i\gamma_{aa'})e^{i\Delta T}}{\omega+\Delta-\omega_{a'c}+i(\gamma_{c}-\gamma_{a'})}.
\end{align}
The delta function arises from the time translation invariance of correlation functions: $\Delta=\omega_{aa'}+i\gamma_{aa'}$ that involves two frequencies of the probe field and the frequency band of the nonequilibrium preparation state $aa'$. Time translational invariance is maintained provided we treat the preparation explicitly via interaction with pulse $\mathcal{E}_1(\omega_1)$ and $\mathcal{E}_1^{*}(\omega_1')$ as in Eqs. (\ref{eq:Siwir}) - (\ref{eq:Siiwir}). This implies that $\omega_1-\omega_1'+\omega+\Delta-\omega=0$. We can then write
\begin{align}
\delta(\omega_1'-\omega_1-\Delta)=\int_{-\infty}^{\infty}d\omega_0\delta(\omega_1'-\omega_1-\omega_0)\delta(\omega_0-\Delta).
\end{align}
The probe pulse by itself does not obey this symmetry as $\Delta\neq 0$. Thus, a description that  excludes the preparation (actinic) pulse $1$ does not have this symmetry. In this case, for a narrowband preparation pulse $\omega_1\simeq\omega_1'$ results in $\Delta=0$ which means that the signal has low frequency resolution limited only by state lifetimes. The preparation pulse launches the vibrational dynamics, which results in high frequency resolution due to joint field plus matter bandwidth as shown in Eqs. (\ref{eq:Siwir1}) - (\ref{eq:Siiwir1}).

When both the pump and the probe pulses are ultrashort, i.e. $\mathcal{E}_1(\tau)=\mathcal{E}_1\delta(\tau)$ is centered at $\tau=0$ and $\mathcal{E}_2(\tau)=\mathcal{E}_2\delta(\tau-T)$ is centered at $\tau=T$, we can neglect the frequency dispersion of the pulse envelopes.
Eqs. (\ref{eq:Siwir}) - (\ref{eq:Siiwir}) then give
\begin{align}\label{eq:Swirf}
&S_{IR}(\omega,T)=-\mathcal{I}\frac{2i}{\hbar^4}\sum_{a,a'}\mu_{ga'}\mu_{ag}^{*}|\mathcal{E}_1|^{2}|\mathcal{E}_2|^2\times\notag\\
&\left[\sum_d\frac{\mu_{a'd}^{*}\mu_{ad}e^{(i\omega_{a'a}-\gamma_{a'a})T}}{\omega-\omega_{ad}+i\gamma_{ad}}+\sum_c\frac{\mu_{a'c}^{*}\mu_{ca}e^{(i\omega_{aa'}-\gamma_{aa'})T}}{\omega-\omega_{a'c}+i\gamma_{a'c}}\right].
\end{align}
One can derive similar SOS expressions for the frequency-gated SRS signals. Eqs. (\ref{eq:Siwir}) - (\ref{eq:Siiwir}) are then recast as
\begin{align}\label{eq:Siwsr}
&S_{SRS}^{(i)}(\omega-\omega_3,T)=\notag\\
&-\mathcal{I}\frac{2i}{\hbar^4}\sum_{a,a',d}\frac{\mu_{ga'}\mu_{ag}^{*}\alpha_{a'd}\alpha_{ad}e^{-(i\omega_{aa'}+\gamma_{aa'})T}}{\omega-\omega_3-\omega_{ad}+i\gamma_{ad}}|\mathcal{E}_3|^2\mathcal{E}_2^{*}(\omega)\notag\\
&\times\mathcal{E}_2(\omega-\omega_{aa'}+i\gamma_{aa'})\mathcal{E}_1^{*}(\omega_{a'}+i\gamma_{a'})\mathcal{E}_1(\omega_a-i\gamma_a),
\end{align}
\begin{align}\label{eq:Siiwsr}
&S_{SRS}^{(ii)}(\omega-\omega_3,T)=\notag\\
&-\mathcal{I}\frac{2i}{\hbar^4}\sum_{a,a',c}\frac{\mu_{ga'}\mu_{ag}^{*}\alpha_{a'c}\alpha_{ac}e^{-(i\omega_{a'a}+\gamma_{a'a})T}}{\omega-\omega_3-\omega_{a'c}+i\gamma_{a'c}}|\mathcal{E}_3|^2\mathcal{E}_2^{*}(\omega)\notag\\
&\times\mathcal{E}_2(\omega-\omega_{a'a}+i\gamma_{a'a})\mathcal{E}_1^{*}(\omega_{a'}+i\gamma_{a'})\mathcal{E}_1(\omega_a-i\gamma_a),
\end{align}

As we did for FDIR, we shall express the signals (\ref{eq:Siwsr}) - (\ref{eq:Siiwsr}) in a form that reveals the broken time translational symmetry. For the general pulse envelope of the pump field $\mathcal{E}_3(\omega_3)$ that enters twice in the signal, e.g. $\mathcal{E}_3^{*}(\omega_3)$ and $\mathcal{E}_3(\omega_3')$, the overall translational symmetry for all six interactions yields
\begin{align}
&\delta(\omega_1'-\omega_1+\omega_3'-\omega_3-\Delta)\notag\\
&=\int_{-\infty}^{\infty}d\omega_0\delta(\omega_1'-\omega_1-\omega_0)\delta(\omega_0+\omega_3'-\omega_3-\Delta),
\end{align}
where the product of two delta functions reveals the broken symmetry for the pump/probe fields when the preparation pulse is excluded. Assuming a narrowband pump $\mathcal{E}_3(t)=\mathcal{E}_3e^{-i\omega_3(t-T)}$ Eqs. (\ref{eq:Siwir1}) - (\ref{eq:Siiwir1})
\begin{align}\label{eq:Siws1}
\tilde{S}_{SRS}^{(i)}&(\omega-\omega_3,T;\Delta)=\notag\\
-&\frac{4\pi i}{\hbar^2}|\mathcal{E}_3|^2\sum_{a,a',d}\frac{\alpha_{a'd}\alpha_{ad}\rho_{aa'}\delta(\Delta-\omega_{a'a}+i\gamma_{a'a})e^{i\Delta T}}{\omega-\omega_3+\Delta-\omega_{a'd}+i(\gamma_{d}-\gamma_{a'})},
\end{align}
\begin{align}\label{eq:Siiws1}
\tilde{S}_{SRS}^{(ii)}&(\omega-\omega_3,T;\Delta)=\notag\\
-&\frac{4\pi i}{\hbar^2}|\mathcal{E}_3|^2\sum_{a,a',c}\frac{\alpha_{a'c}\alpha_{ac}\rho_{a'a}\delta(\Delta-\omega_{aa'}-i\gamma_{aa'})e^{i\Delta T}}{\omega-\omega_3+\Delta-\omega_{a'c}+i(\gamma_{c}-\gamma_{a'})}.
\end{align}

Finally for a broadband probe $\mathcal{E}_2$ the SRS signal (\ref{eq:Siwsr}) - (\ref{eq:Siiwsr}) reduce to
\begin{align}\label{eq:Swsrf}
&S_{SRS}(\omega-\omega_3,T)=-\mathcal{I}\frac{2i}{\hbar^4}\sum_{a,a'}\mu_{ga'}\mu_{ag}^{*}|\mathcal{E}_1|^{2}|\mathcal{E}_2|^2|\mathcal{E}_3|^2\times\notag\\
&\left[\sum_d\frac{\alpha_{a'd}\alpha_{ad}e^{(i\omega_{a'a}-\gamma_{a'a})T}}{\omega-\omega_3-\omega_{ad}+i\gamma_{ad}}+\sum_c\frac{\alpha_{a'c}\alpha_{ca}e^{(i\omega_{aa'}-\gamma_{aa'})T}}{\omega-\omega_3-\omega_{a'c}+i\gamma_{a'c}}\right].
\end{align}

In the SOS protocol  the basis set expansion has to cover the complete vibrational dynamics under investigation, which can be tedious for complex reactive systems and the diagonalization of the resulting Hamiltonian is non-trivial. Model Hamiltonians may be used to truncate the system size and provide an affordable simulation. For example exciton hamiltonians are commonly used to describe multiple excitations in  chromophore aggregates \cite{Abr09}. Once the exact eigenstates are obtained, this protocol allows for the straightforward interpretation of the signals.

\section{Third protocol; Coupling to a classical bath}

A simpler and often more intuitive description can be developed by treating some (bath) degrees of freedom as classical. We start with the ultrafast visible pump and IR probe of the excited vibrational states. We assume that probe pulse is impulsive and set $\mathcal{E}_1(t)=\mathcal{E}_1\delta(t)$, and further evaluate the remaining time integrals using Eqs. (\ref{eq:Ga}) and (\ref{eq:Gc}). The resulting semiclassical $\Delta$ - dispersed signal (\ref{eq:Si}) - (\ref{eq:Sii}) reads

\begin{align}\label{eq:S41}
&\tilde{S}_{IR}(\omega,T;\Delta)=-\frac{2i}{\hbar^4}\int_{-\infty}^{\infty}d\tau_3\int_{\tau_3}^{\infty}dt\times\notag\\
&|\mathcal{E}_1|^2e^{i\omega(t-T)}e^{-i(\omega+\Delta)(\tau_3-T)}\sum_a|\mu_{ag}|^2e^{-2\gamma_at}\times\notag\\
&\left[\sum_c|\mu_{ac}|^2e^{-i\int_{\tau_3}^t\omega_{ac}(t')dt'}+\sum_d|\mu_{ad}|^2e^{i\int_{\tau_3}^t\omega_{ad}(t')dt'}\right],
\end{align}
where $\omega_{\alpha\beta}\equiv|\omega_{\alpha}-\omega_{\beta}|$. Ensemble averaging $\langle...\rangle_e$ over the classical set of trajectories is performed on the signal level $S_{IR}(\omega,T)$. Similarly one can derive the corresponding SRS result when the extra pump pulse is narrow band and can be approximated as $\mathcal{E}_3(t)=\mathcal{E}_3e^{-i\omega_3(t-T)}$. The signal (\ref{eq:Sisr}) - (\ref{eq:Siisr}) then reads

\begin{align}\label{eq:S410}
&\tilde{S}_{SRS}(\omega-\omega_3,T;\Delta)=-\frac{2i}{\hbar^4}\int_{-\infty}^{\infty}d\tau_3\int_{\tau_3}^{\infty}dt\times\notag\\
&|\mathcal{E}_1|^2|\mathcal{E}_3|^2e^{i(\omega-\omega_3)(t-T)}e^{-i(\omega+\Delta)(\tau_3-T)}\sum_a|\mu_{ag}|^2e^{-2\gamma_at}\times\notag\\
&\times\left[\sum_c\alpha_{ac}^2e^{-i\int_{\tau_3}^t\omega_{ac}(t')dt'}+\sum_d\alpha_{ad}^2e^{i\int_{\tau_3}^t\omega_{ad}(t')dt'}\right].
\end{align}

Eqs. (\ref{eq:S41}) and (\ref{eq:S410}) involve a path integral over the stochastic vibrational frequency $\omega_{ac}(t)$ and $\omega_{ad}(t)$. The signal depends not only on the initial and final value of the vibrational frequency $\omega_{\nu\nu'}$, but rather on the entire pathway from time $T$ to the time when the polarization decays to zero. The time dependent frequency $\omega_{\nu,\nu'}(t)$ can be calculated by running classical MD trajectories.  

In the semiclassical protocol of Eqs. (\ref{eq:S41}) - (\ref{eq:S410}) the system is partitioned into a classical bath retaining only the quantum character of a few vibrational modes. For non-reactive systems (i.e. no chemical bonds are broken or formed) evolving on a single adiabatic potential energy surface (i.e. the Born-Oppenheimer approximation remains valid) common molecular dynamics simulations can be used which scale by $N^2$ if all pair-wise electrostatic and van der Waals interactions are  explicitly accounted for. The computational cost can be further  reduced to linear scaling by suitable cutoffs. The quantum character of the vibrations under investigation can be retained by collective solvent coordinates which allow to map the classical dynamics onto \emph{ab initio} derived electrostatic maps \cite{Kwac:JCP:2003,Tomoyuki:JPCA:2005}. 

If the process under investigation is characterized by ultrafast  relaxation in the vicinity of conical intersection as commonly observed in photoreactions the breakdown of  the Born-Oppenheimer approximation requires to treat the system by non-adiabatic on-the-fly molecular dynamics \cite{Barbatti:book:2011}. Based on the independent trajectory approximation the nuclear wavepacket is represented by a swarm of independently evolving trajectories where, within the framework of TullyÕs fewest switches trajectory surface hopping \cite{Tully:JCP:1990,Hammes-Schiffer:JCP:1994}, relaxation between different electronic states is induced by the non-adiabatic coupling (NAC).  Here the numerical effort of the dynamics is shifted to the calculation of excited state gradients and NACs between electronic states on an appropriate quantum chemical level but the construction of global potential energy surfaces is avoided as only the relevant configuration space is explored during the dynamics. The quantum character of vibrations is reconstructed by evaluating the excited state Hessian. The restriction on a few vibrational degrees of freedom allows for an efficient algorithm for the calculation of the semiclassical signal which is based on a mode tracking procedure \cite{Reiher:JCP:2003}, only the desired frequencies and normal mode vectors are obtained. As  the construction of the complete Hessian matrix is avoided linear scaling  with the number of considered vibrational modes can be achieved \cite{Fin13}.

Rather than calculating the path integral numerically we can expand the integral in the exponent into the cumulant series and extract the mean ensemble averaged time dependent frequency $\bar{\omega}_{\nu\nu'}(t)$ and approximate the remaining nuclear motion by harmonic Gaussian fluctuations. The signal calculated in Appendix B may be then expressed in terms of the spectral density of the harmonic bath 
\begin{widetext}
\begin{align}\label{eq:S42}
\tilde{S}_{IR}(\omega,T;\Delta)=&-\frac{2i}{\hbar^4}\int_{-\infty}^{\infty}d\tau_3\int_{\tau_3}^{\infty}dt|\mathcal{E}_1|^2e^{i\omega(t-T)}e^{-i(\omega+\Delta)(\tau_3-T)}\sum_a|\mu_{ag}|^2e^{-2\gamma_at}\notag\\
&\times\left[\sum_c|\mu_{ac}|^2e^{-i\int_{\tau_3}^t\bar{\omega}_{ac}(t')dt'-g_{ac}(T,t)}+\sum_d|\mu_{ad}|^2e^{i\int_{\tau_3}^t\bar{\omega}_{ad}(t')dt'-g_{ad}^{*}(T,t)}\right],
\end{align}
\begin{align}\label{eq:S420}
\tilde{S}_{SRS}(\omega-\omega_3,T;\Delta)=-&\frac{2i}{\hbar^4}\int_{-\infty}^{\infty}d\tau_3\int_{\tau_3}^{\infty}dt|\mathcal{E}_1|^2|\mathcal{E}_3|^2e^{i(\omega-\omega_3)(t-T)}e^{-i(\omega+\Delta)(\tau_3-T)}\sum_a|\mu_{ag}|^2e^{-2\gamma_at}\notag\\
&\times\left[\sum_c\alpha_{ac}^2e^{-i\int_{\tau_3}^t\bar{\omega}_{ac}(t')dt'-g_{ac}(T,t)}+\sum_d\alpha_{ad}^2e^{i\int_{\tau_3}^t\bar{\omega}_{ad}(t')dt'-g_{ad}^{*}(T,t)}\right].
\end{align}
\end{widetext}
where the Gaussian fluctuations  are manifested via the two point linewidth function $g_{aj}^{*}(T,t)$, $j=c,d$ given by 
\begin{align}\label{eq:gac}
g_{ac}(T,t)&=\frac{4\lambda_{ac}T}{\beta\hbar\Lambda}+\left(\frac{2\lambda_{ac}}{\beta\hbar\Lambda^2}-i\frac{\lambda_{ac}}{\Lambda}\right)\notag\\
&\times\left[e^{-\Lambda t}+(\Lambda(t-T)-1)e^{-\Lambda T}\right],
\end{align}
where $\lambda_{ac}$ represents the reorganization energy and $\Lambda$ corresponds to the fluctuation time scale. Note that, the linewidth function depends on both initial and final times, not only the difference. This is a consequence of the non stationary vibrational dynamics.

\section{The genuine temporal and spectral resolution of frequency-gated signals}

The picture emerging from our theory is that the effective temporal and spectral resolution of FDIR and SRS signals is affected by, but not solely controlled by, experimental knobs. Achieving ultrafast resolution requires the active involvement of the entire probe bandwidth. This is eroded when a narrower slice of the pulse is selected by the system.  Below we discuss how the Fourier uncertainty relation between spectral and temporal resolution $\Delta \omega\Delta t>1$ is always satisfied  once $\Delta\omega$ and $\Delta t$ are properly defined. 

The combined spectral and the temporal resolution of these IR and Raman techniques stem from  two interactions with a single device: the probe pulse. Nonlinear  multidimensional  spectroscopy signals depend on several time intervals and there is no conceptual problem in having simultaneous high temporal and spectral resolutions in different independent dimensions \cite{Muk95,Mukamel:PhysRevA.77:2008}. This is not the case  when both dimensions are associated with the same probe pulse. The issue was addressed for Raman detection in \cite{JCP:Mukamel:2011} using a semiclassical treatment of bath coordinates.

Below we present a more general analysis and elaborate on this point for the three protocols and identify the factors that determine the genuine resolution. In the first protocol, the signals (\ref{eq:Si}) - (\ref{eq:Sii}) and (\ref{eq:Sisr}) - (\ref{eq:Siisr}) are given by a sum over paths spanning both branches of the loop. Naively one can argue that a short pulse must interact impulsively with the system at a precisely defined time. This is  not necessarily the case for the following reason: a pulse is a superposition of modes with well defined phases. The broader the bandwidth, the shorter can the pulse be. Eqs. (\ref{eq:Si}) - (\ref{eq:Sii}) and (\ref{eq:Sisr}) - (\ref{eq:Siisr}) show that the relevant range of frequencies that actually contribute to a given signal is spanned by the variable $\Delta$. Thus, only  some of the probe modes contribute to a given signal, and the full bandwidth of the pulse may become immaterial in some cases. A superposition of the relevant modes has a narrow bandwidth and is necessarily less impulsive than the original pulse, thus reducing the temporal resolution. The number of contributing modes is governed by the width of the relevant spectral features of the system and can be easily understood by the selection of the relevant pathways in the joint field plus matter space. Therefore, the resolution is controlled  by the pulse, the measuring device as well as the system in Eqs. (\ref{eq:Si}) - (\ref{eq:Sii}) and (\ref{eq:Sisr}) - (\ref{eq:Siisr}). The relevant range of the $\tau_3$ integration is controlled by the effective bandwidth of $\Delta$, $\Delta=0$ implies a CW probe. In both diagrams ($i$) and ($ii$) the probe is frequency -  dispersed in the detection. If only a single mode is selected for detection one can ask why does the probe duration matter at all? This is apparent from the diagrams  which show that the signal involves two interactions with the probe. Frequency -  dispersed detection only selects the frequency of the last interaction $\mathcal{E}_2^{*}(\omega)$ whereas the other interaction $\mathcal{E}_2(\omega+\Delta)$ can still involve many modes, making the signal depend on the probe bandwidth. The time resolution  is diminished only if the second interaction also selects a single mode so that $\Delta=0$.


Turning now to the second protocol, we first note that the time gated measurements (\ref{eq:Sitir}) - (\ref{eq:Siitir}) and (\ref{eq:Sitsr}) - (\ref{eq:Siitsr}) are given by $\mathcal{E}_2^{*}(t)\mathcal{E}_2(\tau)$ which are peaked around $t=\tau=T$. This means that in a time-gated measurement the signal represents a snapshot of the dynamics taken at fixed time $t$ coming from the flat frequency distribution of the probe pulse. Thus, in the joint field plus matter space, a time-domain measurement selects quantum pathways corresponding to the fixed time measurement that is infinitely broad in frequency. However, the frequency - dispersed signals (\ref{eq:Siwir}) - (\ref{eq:Siiwir}) and (\ref{eq:Siwsr}) - (\ref{eq:Siiwsr}) depend on the product $\mathcal{E}_2^{*}(\omega)\mathcal{E}_2(\omega-\omega_{a'a}+i\gamma_{a'a})$. This creates an uncertainty in the interaction time with the probe which is governed by the vibrational dynamics time scale (spectral width of $\omega_{a'a}$) and bath dephasing $\gamma_{a'a}$. Therefore, the quantum pathways selected by the dynamics of the system yield the effective bandwidth of the probe pulse that interacts with the system. This introduces uncertainty to the interaction time $\tau_3$ in Figs. \ref{fig:IR}b and \ref{fig:Ram}b stemming the finite bath dynamics time scale. The corresponding measurement cannot be viewed as a snapshot of the system, but is determined by the vibrational dynamics that is represented by the coherence between $a$ and $a'$. The bandwidth of the pump pulse which prepares the system in the density matrix $\rho_{aa'}$  is crucial. The energy spread of $\omega_{aa'}$ is controlled by the pump bandwidth and is also a measure of the inverse time scale of the matter dynamics initiated by the pump. If a single state is selected ($\rho_{aa}$) then there is no dynamics and the same signal can be generated by a CW pump tuned generally to level $a$. The pump duration then becomes immaterial. A broad distribution of vibrational states will result in fast dynamics that is affected by the pump duration. The broadband technique amounts to multiple  two-mode experiments in parallel, which is  experimentally convenient since it does not require to scan the frequency, but reveals no additional information beyond the two mode experiment. With initiation, which prepares a wave packet with different $a$, $a'$ pairs, the technique  may be viewed as many  four-wave-mixing (FWM)  experiments done in parallel. This is essentially a broad band FWM which only has  three modes. For comparison CARS is a four-mode process.

SRS that combines a long picosecond pump with a femtosecond probe has low temporal resolution if treated as a 4-field interaction starting with state $a=a'$ similar to Eqs. (\ref{eq:Siwir1}) - (\ref{eq:Siiwir1}) and replacing $\omega\to\omega-\omega_3$. In this case, in order to have a  highly resolved frequency gated signal, the energy conservation law which follows from the time translational invariance enforces $\omega\neq\omega_3$. In the more general case of a broader pump pulse, we have two interactions with $\mathcal{E}_2(\omega_3)$ and $\mathcal{E}_2^{*}(\omega_3')$. The symmetry breaking then involves four modes of the field and bandwidth of the non stationary preparation state: $\Delta-\omega_3+\omega_3'=\omega_{aa'}+i\gamma_{aa'}$.

Finally we turn to the third protocol. It is clear that in a frequency gated measurement the probe pulse bandwidth must be broader than the inverse timescale of the vibrational dynamics. The latter is given by the spread of $\omega_{a'a}$ and the dephasing rate $\gamma_{a'a}$. 
Even if the probe pulse is impulsive and delayed by $T$, the fact that it is an infrared pulse requires to explicitly take into account the pulse shape of $\mathcal{E}_2(\tau)$ and the pulse may not be simply replaced by  $\delta$ function, since it may not be shorter than the infrared period. The optical pulses used in the Raman process in contrast can be shorter than the vibrational period and can be truly impulsive. An infrared pulse can be at most ``semi impulsive'' (i.e. short compared to vibrational relaxation process but not compared to high frequency vibrations $>$300 cm$^{-1}$). In the case of the Raman signal this large bandwidth can be easily realized  for visible frequencies, and the $\delta$-function approximation is well justified. However, this is not as obvious in the case of an IR probe since the bandwidth of the IR pulses are naturally smaller than in the visible range. Therefore, one must keep the probe pulse envelope and the $\delta$-function approximation is not justified.

  \begin{figure*}[t]
\begin{center}
\includegraphics[trim=0cm 0cm 0cm 0cm,angle=0, width=0.95\textwidth]{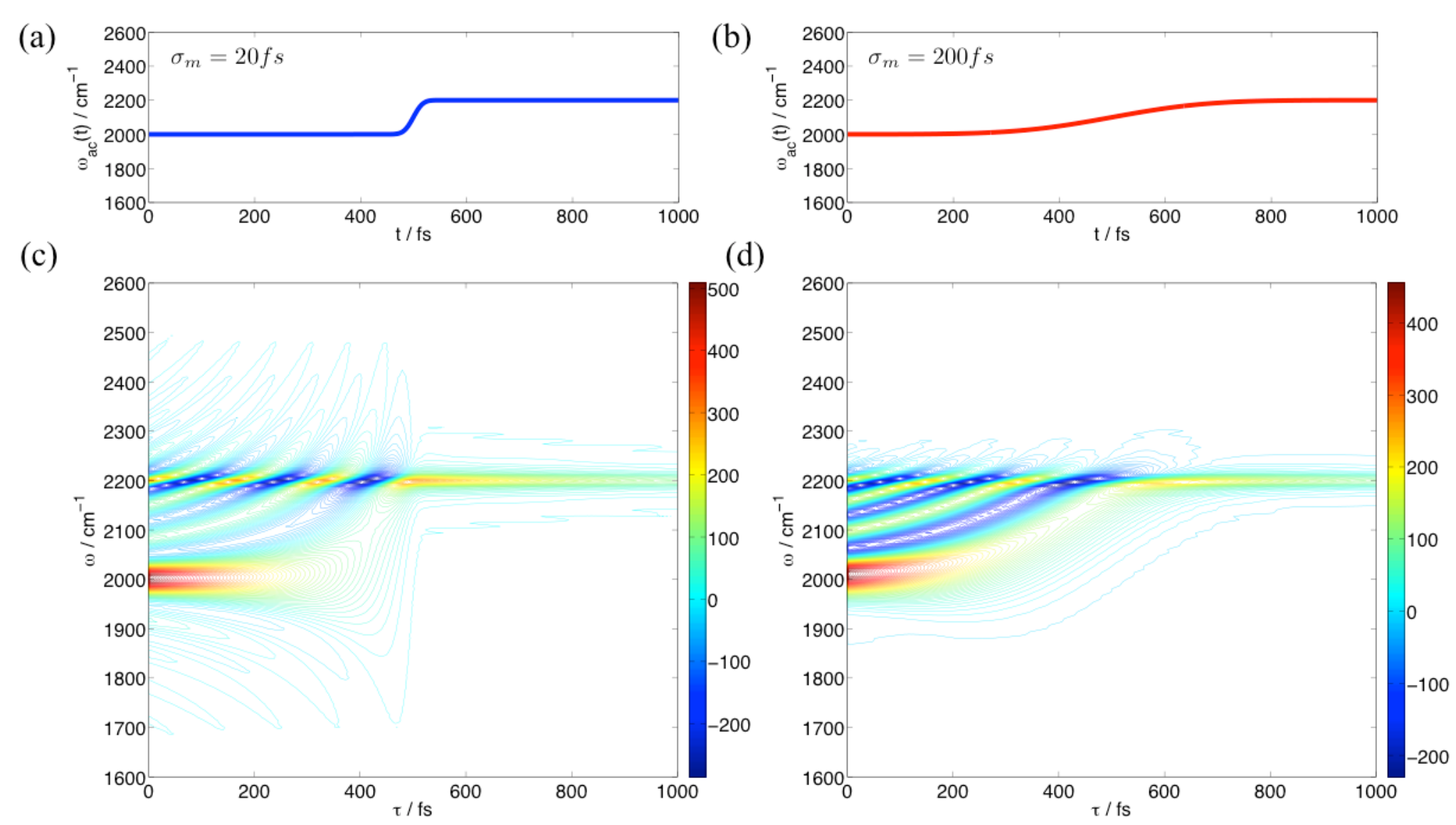}
\end{center}
\caption{(Color online) Frequency profile [Eq. (\ref{eq:err})] - (a) and corresponding 2D the Fourier transform of the $\Delta$ - dispersed signal $\tilde{S}(\omega,T=500~\text{fs},\tau)$ in Eq. (\ref{eq:Swta}) - (c) for $\omega_{ac}^{(0)}=2000$ cm$^{-1}$, $\alpha=200$ cm$^{-1}$, $\sigma_m=20$ fs. (b) and (d) - same as (a) and (c), respectively with $\sigma_m=200$ fs.}
\label{fig:sim}
\end{figure*}

To better illustrate the resolution, we examine the $\Delta$ - dispersed time-domain signal (\ref{eq:Stb}) dressed by the probe pulse
\begin{equation}\label{eq:Std}
\bar{S}(t,T;\Delta)=\int_{-\infty}^{\infty}d\tau\mathcal{E}_2(\tau-T) \tilde{S}(t,T;\tau)e^{i(\omega+\Delta)\tau}.
\end{equation}
and its variation with $\Delta$. For simplicity  in the following we omit the subscript for $\tau_3$. Note that in contrast with the $\tau$ - dispersed signal $\tilde{S}(t,T;\tau)$, $\bar{S}(t,T;\Delta)$ depends on the probe pulse envelope. As discussed above  $\Delta$ may be broadened due to finite timescale of the bath dynamics. The relevant frequency domain  (\ref{eq:Swb}) signal can be calculated using Eq. (\ref{eq:Std})
\begin{align}
S_{IR}(\omega,T)=&\mathcal{I}\int_{-\infty}^{\infty}\frac{d\Delta}{2\pi}\int_{-\infty}^{\infty}d\tau\int_0^{\infty}dte^{i\omega(t-T)-i\Delta\tau}\notag\\
&\times\mathcal{E}_2^{*}(\omega)\bar{S}(t+\tau,T;\Delta).
\end{align}
We first consider a simple example for the bath and calculate the effective bandwidth $\Delta$ within the semiclassical approximation. We assume linear time variation of the matter transition frequency (linear ``matter'' chirp): $\omega_{ac}(t)=\omega_{ac}^{(0)}+\alpha t$, where $\alpha$ is a chirp rate. Taking into account Eq. (\ref{eq:S42}) and assuming a harmonic potential with single states $a$, $c$ and $d$ such that $\omega_{ac}=\omega_{ad}$, setting $\mu_{ad}=\mu_{ac}$ we obtain for the Eq. (\ref{eq:Std})
 \begin{align}
 \bar{S}(t,T;\Delta)&=\theta(t)\int_0^t\mathcal{E}_2(\tau-T)e^{i\Delta\tau-\gamma_a(t+\tau)}\notag\\
 &\times|\mu_{ag}|^2|\mu_{ac}|^2|\mathcal{E}_1|^2\left[e^{i\omega_{ac}^{(0)}(t-\tau)+\frac{i}{2}\alpha(t^2-\tau^2)}+c.c.\right].
  \end{align}
 Assuming a gaussian probe pulse centered around $\tau=T$
 \begin{align}
 \mathcal{E}_2(\tau-T)=\mathcal{E}_2e^{-\frac{(\tau-T)^2}{2\sigma_{pr}^2}-i\omega_0\tau},
 \end{align}
where $\omega_0$ is the central frequency and $\sigma_{pr}$ is the duration of the pulse we obtain
\begin{align}\label{eq:Sdg}
\bar{S}(t,T;\Delta)\sim e^{-\frac{(\Delta-\Delta_0)^2}{2\sigma_{eff}^2}}
 \end{align}
 where $\Delta_0=\omega_0-\omega_{ac}^{(0)}+\alpha(T-\sigma_{pr}^2\gamma_a)$ and $\sigma_{eff}^2=\sigma_{pr}^{-2}+\alpha^2\sigma_{pr}^2$. Note, that effective range for $\Delta$ given by $\sigma_{eff}$ contains two contributions.  One is the inverse duration of the pulse, and the second is governed by  $\alpha$ - a characteristic timescale of the matter dynamics. This effect is similar to the broadening of a chirped pulse compared to the transform-limited pulse with the chirp added by the matter, instead.

\begin{figure*}[t]
\begin{center}
\includegraphics[trim=0cm 0cm 0cm 0cm,angle=0, width=0.95\textwidth]{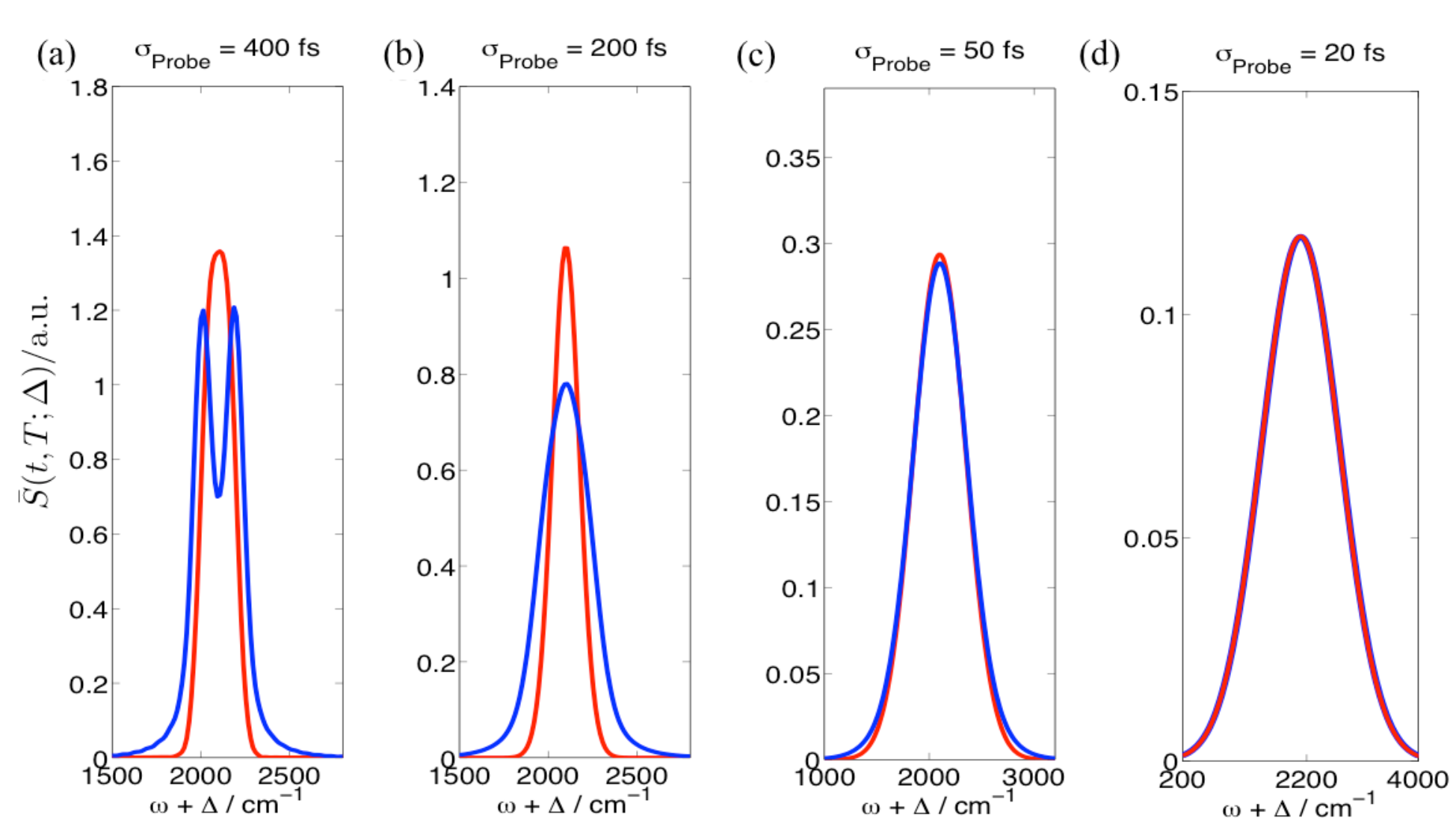}
\end{center}
\caption{(Color online)  The $\Delta$ - dispersed signal Eq. (\ref{eq:Std}) vs $\Delta$ for fast ($\sigma_m=20$ fs, blue) and slow ( $\sigma_m=200$ fs, red) switchover of the vibrational frequency as depicted in Fig. \ref{fig:sim} (a) and (b), respectively. We assume resonant excitation $\omega_0=\omega_{ac}^{(0)}$. Various panels represent different values of the pulse duration $\sigma_{pr}=400$ fs - (a), 200 fs - (b), 50 fs - (c), and 20 fs - (d).}
\label{fig:sim1}
\end{figure*}

We would like to capture the matter dynamics at a given time scale $\alpha^{-1}$. For a long pulse the dominant contribution  to $\sigma_{eff}$  comes from the matter which ensures high frequency resolution. In the limit of resonant CW excitation ($\omega_0=\omega_{ac}^{(0)}$) Eqs. (\ref{eq:Sdg}) gives $\delta(\Delta)$. The latter implies that the original $\Delta$ - dispersed signal (\ref{eq:S42}) has no time resolution with respect to $\tau$. This result is independent of time delay of the probe pulse $T$. In the opposite limit when the pulse duration is small the leading contribution comes from the pulse and $\sigma_{eff}\simeq\sigma_{pr}^{-1}$.  Therefore, high temporal resolution is accompanied by poor spectral resolution and vice versa. In both limits, the time and frequency resolution are not independent or solely controlled by external manipulation of pulse parameters. Rather they are governed by a combination of pulse and matter parameters.
This simple example provides some basic intuition. However in this linear matter chirp model the transition frequency is changing in an unbounded fashion. We next consider a more realistic model where the transition frequency switches between two values during a finite time interval. For instance
 \begin{align}\label{eq:err}
 \omega_{ac}(t)=\omega_{ac}^{(0)}+\frac{1}{2}\alpha\left[\mathcal{F}\left(\frac{t_0}{\sigma_m}\right)-\mathcal{F}\left(\frac{t_0-t}{\sigma_m}\right)\right],
 \end{align}
where $\mathcal{F}(t)=\frac{2}{\sqrt{\pi}}\int_0^tdxe^{-x^2}$ is the error function. The transition matter frequency switches from its initial $\omega_{ac}^{(0)}$ to its final value $\omega_{ac}^{(0)}+\alpha$ during time interval $\sigma_m$ in the vicinity of $t_0$. Fig. \ref{fig:sim}a depicts Eq. (\ref{eq:err})  with $\omega_{ac}^{(0)}=2000$ cm$^{-1}$, $\alpha=200$ cm$^{-1}$, $\sigma_m$=20 fs, $t_0$=500 fs. Fig. \ref{fig:sim}c shows the Fourier transform of the $\Delta$ - dispersed signal
\begin{align}\label{eq:Swta}
\tilde{S}(\omega,T;\tau)=\int_{-\infty}^{\infty}\frac{d\Delta}{2\pi}\tilde{S}(\omega,T;\Delta)e^{-i(\omega+\Delta)\tau}
\end{align}
vs $\omega$ and $\tau$. For $T=500$ fs one can see how the pattern evolves with dominating emission peak at initial frequency $\omega_{ac}^{(0)}=2000$ $cm^{-1}$ for short times $\tau<500$ fs turning into the final frequency $\omega_{ac}^{(0)}+\alpha=2200$ cm$^{-1}$ for longer times $\tau>500$ fs. The oscillatory region of the plot  for times shorter than $\gamma_a^{-1}=1$ ps shows the frequency beating and matter chirp. At longer times the signal decays exponentially $\sim e^{-\gamma_a\tau}$. For slower dynamics, $\sigma_m=200$ fs (see Fig. \ref{fig:sim}b) the $\Delta$  - dispersed signal (\ref{eq:Swta}) plotted in Fig. \ref{fig:sim}d is similar to Fig. \ref{fig:sim}c but is stretched according to the longer time scale $\sigma_m$.

In order to determine the time and frequency resolution for the system dynamics given by Eq. (\ref{eq:err}) with $\sigma_m=20$ fs and $\sigma_m=200$ fs, we plotted the $\Delta$ - dispersed  time-domain signal (\ref{eq:Std}) for different values of the probe pulse duration $\sigma_{pr}$ in Fig. \ref{fig:sim1}a-d. For long probe $\sigma_{pr}=400$ fs Fig. \ref{fig:sim1}a shows that the slow matter dynamics results in a single peak at the final frequency $\omega+\Delta=\omega_{ac}^{(0)}+\alpha$, whereas fast dynamics gives two peaks which correspond to the initial and final frequencies. 
Further increase of the pulse duration [not shown] does not change the fast dynamics case while for slow dynamics the two peaks become narrower. It means that the high frequency resolution is accompanied by poor time resolution in this case. 
For a shorter pulse $\sigma_{pr}=200$ fs both fast and slow dynamics give a single emission peak  centered at final frequency $\omega_{ac}^{(0)}+\alpha$. However fast dynamics yields a larger bandwidth due to  combined pulse and matter bandwidths (see Fig. \ref{fig:sim1}b). Further decrease of the pulse duration  for $\sigma_{pr}=50$ (Fig. \ref{fig:sim1}c) and 20 fs (Fig. \ref{fig:sim1}d) shows that the fast dynamics converges and becomes indistinguishable from the slow dynamics at $\sigma_{pr}\simeq\sigma_m=20$ fs (Fig. \ref{fig:sim1}d). In this case the spectrum does not carry any matter information and looses its frequency resolution. It simply gives the Fourier transform of the probe pulse. Therefore, for long pulse, the spectrum has perfect frequency and poor time resolution. This corresponds to the CW experiment, when the initiation pulse prepares the system in equilibrium population state described by $\rho_{aa}$. In this case time translation invariance via (\ref{eq:bdf}) yields $\omega_1=\omega_1'$ and consequently $\omega=\omega'$. In the opposite limit of the short pulse, the perfect time resolution is accompanied by poor frequency resolution. The resulting spectrum will not contain any relevant matter information and will be given by a Fourier transform of the probe pulse. In both limits the time and the frequency resolution are not independent and are governed by a combination of matter and field parameters.

\section{Discussion}

We have employed a superoperator diagrammatic techniques to derive similar expressions  for stimulated signals  detected by frequency  - dispersed transmission of  a broadband  IR probe and stimulated Raman signals following the broadband visible pump pulse. The resolution is determined by both field and matter degrees of freedom and cannot be solely controlled by the experimental apparatus. The time and frequency resolution was analyzed using three representations and the corresponding computational protocols for the signal. Loop diagrams  provide a convenient  compact tool for computing and interpreting these signals in terms of the evolving vibronic wavepacket. The vibrational resonances are generated during a single time interval in this diagram where the wave function propagates backward from the observation time corresponding to the ket interaction with the probe field  and further to the previous bra interaction with the same probe which is close to the delay time $T$ relative to the preparation field. Note that $t-T$ is a time interval between two successive interactions along the loop but not in real time. A completely time ordered description based on ladder diagrams will separate the loop into several terms \cite{Rah10}. 

The SRS signals may be obtained from the FDIR expressions by the substitution $V_n\to\alpha_n$ and $\mathcal{E}_2(t)\to\mathcal{E}_2(t)\mathcal{E}_3^{*}(t)$. Note that due to the additional narrow band field $3$ in frequency domain SRS the infrared frequency $\omega$ must be replaced by the Raman shift $\omega-\omega_3$. Both SRS and FDIR are given by two diagrams which represent different physical processes. Diagrams ($i$) and ($ii$) in Fig. \ref{fig:IR}b correspond to emission and absorption, respectively of the IR probe pulse, whereas in the case of Fig. \ref{fig:Ram}b each diagram contains both emission and absorption (Stokes and anti-Stokes process). In FDIR we can make the rotating wave approximation (RWA) for the radiation matter coupling and obtain Eq. (\ref{eq:H1}) where $V$, $V^{\dagger}$ are non Hermitian operators. The RWA does not apply for off resonant Raman where we have for the Raman part of Eq. (\ref{eq:Hp})
\begin{equation}
H'(t)=\alpha[\mathcal{E}_2^{*}(t)\mathcal{E}_3(t)+\mathcal{E}_2(t)\mathcal{E}_3^{*}(t)],
\end{equation}
where $\alpha$ is a real (Hermitian) operator $\alpha=\alpha^{\dagger}$. Both $\mathcal{E}_2^{*}(t)\mathcal{E}_3(t)$ and $\mathcal{E}_2(t)\mathcal{E}_3^{*}(t)$ can excite or de-excite the vibrations (Stokes and anti-Stokes process) as permitted by their bandwidths whereas in the FDIR case (Eq. (\ref{eq:H1})) $\mathcal{E}_2$ excites and $\mathcal{E}_2^{\dagger}$ de-excite the vibrations. This is an important distinction, especially in the case of CARS signals (diagram $ii$ in Fig. \ref{fig:Ram}b), which involve four field modes with different wave vectors where the spatial phase matching becomes crucial \cite{Dor11}.


\acknowledgements

We gratefully acknowledge the support of the National Institute of Health Grant No. GM-59230, National Science Foundation
through Grant No. CHE-1058791 and computations are supported by CHE-0840513,  the Chemical Sciences, Geosciences and
Biosciences Division, Office of Basic Energy Sciences, Office of Science and US Department of Energy. B. P. F. gratefully acknowledges support from the Alexander-von-Humboldt
Foundation through the Feodor-Lynen program.

\appendix

\section{Time-gated signals}\label{app:time}

Below we present the time gated signals corresponding to the frequency gated expressions given in the main text. We first read off the FDIR signal from the diagrams  similar to Eqs. (\ref{eq:Si}) - (\ref{eq:Sii}) and introduce the $\tau$ - dispersed signal in time domain analogues to Eq. (\ref{eq:Swb})
\begin{align}\label{eq:Stb}
S_{IR}(t,T)=\mathcal{I}\int_{-\infty}^{t}d\tau\mathcal{E}_2^{*}(t-T)\mathcal{E}_2(\tau-T)\tilde{S}_{IR}(t,T;\tau),
\end{align} 
where $\tilde{S}_{IR}(t,T;\tau)=\tilde{S}_{IR}^{(i)}(t,T;\tau)+\tilde{S}_{IR}^{(ii)}(t,T;\tau)$ where
\begin{align}\label{eq:PB}
\tilde{S}_{IR}^{(i)}(t,T;\tau)&=\frac{2}{\hbar}\int_{-\infty}^td\tau_1\int_{-\infty}^{\tau}d\tau_5\mathcal{E}_1^{*}(\tau_5)\mathcal{E}_1(\tau_1)\notag\\
&\times\langle V_eG^{\dagger}(\tau,\tau_5)V_n^{\dagger}G^{\dagger}(t,\tau)V_nG(t,\tau_1)V_e^{\dagger}\rangle,
\end{align}
\begin{align}\label{eq:PC}
\tilde{S}_{IR}^{(ii)}(t,T;\tau)&=\frac{2}{\hbar}\int_{-\infty}^td\tau_1\int_{-\infty}^{\tau}d\tau_5\mathcal{E}_1(\tau_5)\mathcal{E}_1^{*}(\tau_1)\notag\\
&\times\langle V_eG^{\dagger}(t,\tau_1)V_nG(t,\tau)V_n^{\dagger}G(\tau,\tau_5)V_e^{\dagger}\rangle.
\end{align}
$\tilde{S}(t,T;\tau)$ is the signal at time $t$ resulting from  interaction with $\mathcal{E}_2$ at time $\tau-T$. The signal is obtained by integration over $\tau$. The corresponding SRS signal reads

\begin{align}\label{eq:Pisr}
&\tilde{S}_{SRS}^{(i)}(t,T;\tau)=\frac{2}{\hbar}\int_{-\infty}^td\tau_1\int_{-\infty}^{\tau_3}d\tau_5\mathcal{E}_1^{*}(\tau_5)\mathcal{E}_1(\tau_1)\times\notag\\
&\mathcal{E}_3(t-T)\mathcal{E}_3^{*}(\tau-T)\langle V_eG^{\dagger}(\tau,\tau_5)\alpha_nG^{\dagger}(t,\tau)\alpha_nG(t,\tau_1)V_e^{\dagger}\rangle,
\end{align}
\begin{align}\label{eq:Piisr}
&\tilde{S}_{SRS}^{(ii)}(t,,T;\tau)=\frac{2}{\hbar}\int_{-\infty}^td\tau_1\int_{-\infty}^{\tau_3}d\tau_5\mathcal{E}_1(\tau_5)\mathcal{E}_1^{*}(\tau_1)\times\notag\\
&\mathcal{E}_3(t-T)\mathcal{E}_3^{*}(\tau-T)\langle V_eG^{\dagger}(t,\tau_1)\alpha_nG(t,\tau)\alpha_nG(\tau,\tau_5)V_e^{\dagger}\rangle.
\end{align}
Eqs. (\ref{eq:Pisr}) - (\ref{eq:Piisr}) are analogue of (\ref{eq:Sisr}) - (\ref{eq:Siisr}).

The time-domain FDIR signals Eqs. (\ref{eq:PB}) - (\ref{eq:PC}) can be recast using SOS expansion
\begin{align}\label{eq:Sitir}
&\tilde{S}_{IR}^{(i)}(t,T;\tau)=\frac{2}{\hbar}\theta(\tau)\theta(t)\sum_{a,a',d}\mu_{ga'}\mu_{ag}^{*}\mu_{a'd}^{*}\mu_{ad}\times\notag\\
&\mathcal{E}_1^{*}(\omega_{a'}+i\gamma_{a'})\mathcal{E}_1(\omega_a-i\gamma_a)e^{-(i\omega_{ad}+\gamma_{ad})t+(i\omega_{a'd}+\gamma_d-\gamma_{a'})\tau}.
\end{align}
\begin{align}\label{eq:Siitir}
&\tilde{S}_{IR}^{(ii)}(t,T;\tau)=-\frac{2}{\hbar}\theta(\tau)\theta(t)\sum_{a,a',c}\mu_{ga'}\mu_{ag}^{*}\mu_{a'c}^{*}\mu_{ca}\times\notag\\
&\mathcal{E}_1^{*}(\omega_{a'}+i\gamma_{a'})\mathcal{E}_1(\omega_a-i\gamma_a)e^{-(i\omega_{a'c}+\gamma_{a'c})t+(i\omega_{ac}+\gamma_c-\gamma_{a})\tau},
\end{align}
that are analogues to Eqs. (\ref{eq:Siwir}) - (\ref{eq:Siiwir}). The corresponding SRS signal reads

\begin{align}\label{eq:Sitsr}
S_{SRS}^{(i)}&(t,T)=\mathcal{I}\frac{2i}{\hbar^4}\theta(\tau)\theta(t)\sum_{a,a',d}\mu_{ga'}\mu_{ag}^{*}\alpha_{a'd}\alpha_{ad}\notag\\
&\times\mathcal{E}_1^{*}(\omega_{a'}+i\gamma_{a'})\mathcal{E}_1(\omega_a-i\gamma_a)|\mathcal{E}_3|^2\notag\\
&\times e^{-i\omega_3(t-\tau)-(i\omega_{ad}+\gamma_{ad})t+(i\omega_{a'd}+\gamma_d-\gamma_{a'})\tau},
\end{align}
\begin{align}\label{eq:Siitsr}
S_{SRS}^{(ii)}&(t,T)=-\mathcal{I}\frac{2i}{\hbar^4}\theta(\tau)\theta(t)\sum_{a,a',c}\mu_{ga'}\mu_{ag}^{*}\alpha_{a'c}\alpha_{ca}\notag\\
&\times\mathcal{E}_1^{*}(\omega_{a'}+i\gamma_{a'})\mathcal{E}_1(\omega_a-i\gamma_a)|\mathcal{E}_3|^2\notag\\
&\times e^{-i\omega_3(t-\tau)-(i\omega_{a'c}+\gamma_{a'c})t+(i\omega_{ac}+\gamma_c-\gamma_{a})\tau},
\end{align}

For the ultrafast probe $\mathcal{E}_2(t-T)=\mathcal{E}_2\delta(t-T)$ the $\tau$ - dispersed signal Eq. (\ref{eq:Sitir}) - (\ref{eq:Siitir}) then results in the full signal (\ref{eq:Stb})
\begin{align}\label{eq:Stirf}
&S_{IR}(t,T)=\mathcal{I}\frac{2i}{\hbar^4}\theta(t)\delta(t-T)\sum_{a,a'}\mu_{ga'}\mu_{ag}^{*}|\mathcal{E}_1|^2|\mathcal{E}_2|^2\notag\\
&\times\left[\sum_d\mu_{a'd}^{*}\mu_{ad}e^{(i\omega_{a'a}-\gamma_{a'a})T}-\sum_c\mu_{a'c}^{*}\mu_{ca}e^{(i\omega_{aa'}-\gamma_{aa'})T}\right].
\end{align}
The corresponding SRS signal reads
\begin{align}\label{eq:Stirf}
&S_{SRS}(t,T)=\mathcal{I}\frac{2i}{\hbar^4}\theta(t)\delta(t-T)\sum_{a,a'}\mu_{ga'}\mu_{ag}^{*}|\mathcal{E}_1|^2|\mathcal{E}_2|^2|\mathcal{E}_3|^2\notag\\
&\times\left[\sum_d\alpha_{a'd}\alpha_{ad}e^{(i\omega_{a'a}-\gamma_{a'a})T}-\sum_c\alpha_{a'c}\alpha_{ca}e^{(i\omega_{aa'}-\gamma_{aa'})T}\right].
\end{align}

\section{Coupling to a classical bath}\label{app:sc}

We assume that the system is coupled to a harmonic bath. The molecule is represented by the Hamiltonian
\begin{equation}\label{eq:H0}
H=\sum_{\alpha=g,b}|\alpha\rangle H_{\alpha} \langle\alpha |+|a\rangle H_a(\mathbf{q})\langle a|+|c\rangle H_c(\mathbf{q})\langle c|,
\end{equation}  
where $H_{\beta}(\mathbf{q})$, $\beta=a,c$ is an operator in the nuclear Hilbert space, that is given by
\begin{equation}\label{eq:Ha1}
H_a(\mathbf{q})=\sum_j\left[\frac{\tilde{p}_j^2}{2m_j}+\frac{1}{2}m_j\omega_j^2(\tilde{q}_j)\tilde{q}_j^2\right],
\end{equation}
\begin{equation}\label{eq:Hc1}
H_c(\mathbf{q})=\hbar\omega_{ac}^{(0)}+\sum_j\left[\frac{\tilde{p}_j^2}{2m_j}+\frac{1}{2}m_j\omega_j^2(\tilde{q}_j)(\tilde{q}_j+\tilde{d}_j)^2\right],
\end{equation}
where $\omega_j(\tilde{q}_j)$ represents the time dependent frequency profile of the isomerization process. Introducing the dimensionless coordinate $q_j=(m_j\omega_j/\hbar)^{1/2}\tilde{q}_j$, displacement $d_j=(m_j\omega_j/\hbar)^{1/2}\tilde{d}_j$ and momentum $p_j=(m_j\omega_j\hbar)^{1/2}\tilde{p}_j$, Eqs. (\ref{eq:Ha1}) - (\ref{eq:Hc1}) read
\begin{equation}
H_a(\mathbf{q})=\frac{1}{2}\sum_j\hbar\omega_j[p_j^2+q_j^2],
\end{equation}
\begin{equation}
H_c(\mathbf{q})=\hbar\omega_{ac}^{(0)}+\frac{1}{2}\sum_j\hbar\omega_j[p_j^2+(q_j+d_j)^2].
\end{equation}
We next define the vibrational frequency $\hbar\omega_{ac}=\hbar\omega_{ac}^{(0)}+\frac{1}{2}\sum_jd_j^2\omega_j(q_j)$ and potential energy 
\begin{equation}
U_{ac}=H_c-H_a-\hbar\omega_{ac}=\hbar\sum_j\omega_j(q_j)d_jq_j.
\end{equation}
The dipole operator is given by
\begin{equation}
V=\sum_{\alpha,\alpha'}\mu_{\alpha\alpha'}|\alpha\rangle\langle \alpha'|,
\end{equation}
where the summation runs over $\alpha,\alpha'=g,a,c,b$, and $\alpha\neq\alpha'$. Note that at this point we neglect any nuclear of the dipole operators $\mu_{\alpha\alpha'}$ (Condon approximation).

Following the definition of the frequency dispersed signal (\ref{eq:Sw}) we note that the angular brackets $\langle ...\rangle$ in (\ref{eq:Sw}) now represent the average over the bath degrees of freedom. Nuclear dynamics can be approximated by a combination of classical dynamics and additional phases. Introducing the reference Hamiltonian \cite{Rah101}
\begin{equation}
H_{ref}(\tau) = \begin{cases} H_g, & \mbox{if } \tau<\tau_1,\tau_5, \\ H_a, & \mbox{if } \tau\geq\tau_1,\tau_5 \end{cases}
\end{equation}
The Green's function can then be recast with respect to the reference Hamiltonian
\begin{align}
G_{\alpha}(t_1,t_2)=\theta(t_1-t_2)&\exp_+\left[-\frac{i}{\hbar}\int_{t_2}^{t_1}d\tau H_{ref}(\tau)\right]\notag\\
\times&\exp_+\left[-\frac{i}{\hbar}\int_{t_2}^{t_1}d\tau U_{\alpha}(\tau)\right],
\end{align}
where the $``+"$ subscript correspond to the positive time ordering. We assume in Eq. (\ref{eq:H0}) the nuclear dynamics occurs only in the singly excited manifold (states $a$ and $c$). Therefore for $\alpha=a,b$
\begin{equation}\label{eq:Ga}
G_{\alpha}(t_1,t_2)=\theta(t_1-t_2)e^{-(i\omega_{\alpha}+\gamma_{\alpha})(t_1-t_2)},
\end{equation}
while 
\begin{equation}\label{eq:Gc}
G_{c}^{\dagger}(t,\tau_3)=\theta(t-\tau_3)e^{i\omega_a(t-\tau_3)}\exp_-\left[\frac{i}{\hbar}\int_{\tau_3}^td\tau U_{ac}(\tau)\right],
\end{equation}
where 
\begin{equation}
U_{ac}(\tau)=e^{\frac{i}{\hbar}H_a\tau}[H_c-H_a-\hbar\omega_{ac}]e^{-\frac{i}{\hbar}H_a\tau}.
\end{equation}
Substituting this in Eqs. (\ref{eq:PB}) - (\ref{eq:PC}) and (\ref{eq:Pisr}) - (\ref{eq:Piisr}) we then get Eqs. (\ref{eq:S41}) and (\ref{eq:S410}), respectively.


In the reduced description when we treat bath degrees of freedom separately signals (\ref{eq:Si}) - (\ref{eq:Sii}) and (\ref{eq:Sisr}) - (\ref{eq:Siisr})  contain in principle two averaging operations. First is averaging over statistical ensemble of classical trajectories $\langle...\rangle_e$. For a fixed trajectory one has to evaluate the average over the bath degrees of freedom $\langle ...\rangle_b$.  In order to evaluate the correlation function one has to consider the microscopic stochastic dynamics of the nuclei. For a fixed trajectory we evaluate the bath averaging $\langle U_{\nu\nu'}(\tau)\rangle_b=\hbar\omega_{\nu\nu'}(\tau)$ and obtain Eqs. (\ref{eq:S41}) - (\ref{eq:S410}).  We then note that the frequency averaging over trajectories  $\langle \omega_{\nu\nu'}(\tau)\rangle_e=\bar{\omega}_{\nu\nu'}$. One can further add a harmonic fluctuations around the mean value $\bar{\omega}_{\nu\nu'}$ via cumulant expansion. Note that for gaussian fluctuations this expansion is same for all trajectories. We thus obtain
\begin{align}\label{eq:Ue}
&\Bigg\langle\Bigg\langle\exp_-\left(\frac{i}{\hbar}\int_T^td\tau U_{ac}(\tau)\right)\rho_g\Bigg\rangle_b\Bigg\rangle_e=e^{i\int_{T}^t\bar{\omega}_{ac}(\tau)d\tau}\times\notag\\
&\left[1+\mathcal{T}_{-}\left(\frac{i}{\hbar}\right)^2\int_T^td\tau_1\int_T^{\tau_1}d\tau_2\langle U_{ac}(\tau_1)U_{ac}(\tau_2)\rho_g\rangle_b+...\right] 
\end{align}
Note that the linear term in expansion (\ref{eq:Ue}) does not depend on time $\langle U_{ac}(\tau)\rho_g\rangle=\langle U_{ac}\rho_g(\tau)\rangle=\langle U_{ac}\rho_g(0)\rangle$. We further obtain the cumulant expansion by postulating that expansion (\ref{eq:Ue}) can be written as exponentiated in terms of power of $U_{ac}$. Introducing the two-time linewidth function
\begin{equation}
g_{ac}(t_1,t_2)=\int_{t_1}^{t_2}d\tau_1\int_{t_1}^{\tau_1}d\tau_2C_{ac}(\tau_2),
\end{equation}
where $C_{ac}(\tau_2)=\hbar^{-2}\langle U_{ac}(\tau_2)U_{ac}(0)\rho_g\rangle$ represents the spectral density that contains all the microscopic information necessary for calculating the optical response functions within the second order cumulant approximation. We first note that $C(-t)=C{*}(t)$. We next separate it into real and imaginary part $C(t)=C'(t)+C''(t)$. Using the fluctuation-dissipation and detailed balance theorem one may show that
\begin{equation}
\tilde{C}(\omega)=[1+\coth(\beta\hbar\omega/2)]\tilde{C}''(\omega),
\end{equation}
where $\tilde{C}(\omega)=\int_{-\infty}^{\infty}dte^{i\omega t}C(t)$ and $\beta=1/k_BT_a$ with the ambient temperature $T_a$ and Boltzmann constant $k_B$. For the continuous spectrum of harmonic fluctuations one can use the overdamped Brownian oscillator model, i.e.
\begin{equation}
\tilde{C}''(\omega)=2\lambda\frac{\omega\Lambda}{\omega^2+\Lambda^2},
\end{equation}
where $\lambda$ represents the reorganization energy and $\Lambda$ corresponds to the fluctuation time scale. In this case the linewidth function  is given by  Eq. (\ref{eq:gac}).

%

\end{document}